\documentclass[twocolumn,trackchanges]{aastex631}

\usepackage{amsmath}

\begin{document}

\title{PANCAKE: Python bAsed Numerical Color-magnitude-diagram Analysis pacKagE}

\correspondingauthor{Zheng Zheng, Yun Zheng}
\email{zz@bao.ac.cn, yunzh168@gmail.com}

\author[0009-0002-9472-3033]{Yun Zheng}
\affiliation{Beijing 21st Century School, Beijing 100142, People's Republic of China}
\affiliation{Zhejiang Lab, Hangzhou, Zhejiang 311121, People's Republic of China}
\affiliation{Kavli Institute for Astronomy and Astrophysics, Peking University, Beijing 100871, People's Republic of China}

\author[0000-0002-3180-2327]{Yujiao Yang}
\affiliation{School of Astronomy and Space Science, University of Chinese Academy of Sciences, Beijing 100049, People's Republic of China}

\author[0000-0002-8744-3546]{Yong-Kun Zhang}
\affiliation{National Astronomical Observatories, Chinese Academy of Sciences, Beijing 100101, People's Republic of China}

\author[0009-0005-9546-4573]{Zheng Zheng}
\affiliation{National Astronomical Observatories, Chinese Academy of Sciences, Beijing 100101, People's Republic of China}
\affiliation{Key Laboratory of Radio Astronomy and Technology, Chinese Academy of Sciences, Beijing, 100101, People's Republic of China}

\author[0000-0002-6593-8820]{Jing Wang}
\affiliation{Kavli Institute for Astronomy and Astrophysics, Peking University, Beijing 100871, People's Republic of China}

\author[0000-0002-8057-0294]{Lister Staveley-Smith}
\affiliation{International Centre for Radio Astronomy Research, University of Western Australia, 35 Stirling Highway, Crawley, WA 6009, Australia}
\affiliation{ARC Centre of Excellence for All-Sky Astrophysics in 3 Dimensions (ASTRO 3D), Australia}

\author[0000-0002-9390-9672]{Chao-Wei Tsai}
\affiliation{National Astronomical Observatories, Chinese Academy of Sciences, Beijing 100101, People's Republic of China}
\affiliation{Institute for Frontiers in Astronomy and Astrophysics, Beijing Normal University, Beijing 102206, People's Republic of China}
\affiliation{University of Chinese Academy of Sciences, Beijing 100049, People's Republic of China}

\author[0000-0003-3010-7661]{Di Li}
\affiliation{New Cornerstone Science Laboratory, Department of Astronomy, Tsinghua University, Beijing 100084, People's Republic of China}
\affiliation{National Astronomical Observatories, Chinese Academy of Sciences, Beijing 100101, People's Republic of China}
\affiliation{Zhejiang Lab, Hangzhou, Zhejiang 311121, People's Republic of China}

\author[0000-0002-1802-6917]{Chao Liu}
\affiliation{National Astronomical Observatories, Chinese Academy of Sciences, Beijing 100101, People's Republic of China}
\affiliation{University of Chinese Academy of Sciences, Beijing 100049, People's Republic of China}

\author{Jingjing Hu}
\affiliation{Zhejiang Lab, Hangzhou, Zhejiang 311121, People's Republic of China}

\author[0009-0000-6108-2730]{Huaxi Chen}
\affiliation{Zhejiang Lab, Hangzhou, Zhejiang 311121, People's Republic of China}

\author[0000-0003-4811-2581]{Donghui Quan}
\affiliation{Zhejiang Lab, Hangzhou, Zhejiang 311121, People's Republic of China}

\author[0009-0009-2303-395X]{Yinghui Zheng}
\affiliation{National Astronomical Observatories, Chinese Academy of Sciences, Beijing 100101, People's Republic of China}
\affiliation{University of Chinese Academy of Sciences, Beijing 100049, People's Republic of China}

\author{Hangyuan Li}
\affiliation{Kavli Institute for Astronomy and Astrophysics, Peking University, Beijing 100871, People's Republic of China}

\begin{abstract}
Stellar populations serve as a fossil record of galaxy formation and evolution, providing crucial information about the history of star formation and galaxy evolution. The color-magnitude diagram (CMD) stands out as the most accurate tool currently available for inferring the star formation histories (SFHs) of nearby galaxies with stellar-resolved multiband data. The launch of new space telescopes, including JWST, EUCLID, and the upcoming CSST and Roman, will significantly increase the number of stellar-resolved galaxies over the next decade. A user-friendly and customizable CMD fitting package would be valuable for galaxy evolution studies with these data. We develop an open-source Python-based package named \textsc{pancake}, which is fast and accurate in determining SFHs and stellar population parameters in nearby galaxies. We have validated our method via a series of comprehensive tests. First, \textsc{pancake} performs well on mock data, meanwhile the random and systematic uncertainties are quantified. Second, \textsc{pancake} performs well on observational data containing a star cluster and 38 dwarf galaxies (50 fields). Third, the star formation rate (SFR) from \textsc{pancake} is consistent with the SFR from FUV photometry. To ensure compatibility and accuracy, we have included isochrone libraries generated using PARSEC for most of the optical and near-infrared filters used in space telescopes such as HST, JWST, and the upcoming CSST. 
\end{abstract}


\section{Introduction} 
The star formation history (SFH) is pivotal in understanding the evolutionary processes of galaxies. Color-magnitude diagrams (CMDs) serve as an important tool for studying the SFHs of nearby galaxies \citep{tosi1991star, cole1999stellar, momany2005hst, mcconnachie2006stellar}. These CMD studies have analyzed the early star formation histories of nearby galaxies to unravel whether they are remnants of the reionization era or more recently formed \citep{monelli2010acs3, skillman2017islands, bettinelli2018star}. Furthermore, these CMD studies have revealed the multifaceted processes that govern the SFHs and the evolution of nearby galaxies, including external factors such as tidal interactions and mergers, as well as internal mechanisms such as feedback \citep{mcquinn2010nature, weisz2014star1, weisz2014star2, weisz2015star, gallart2015acs, skillman2017islands}. 

The CMD-based SFH measurement method has been developed over three decades. \cite{tosi1991star} utilized the luminosity functions within CMDs and initiated a qualitative comparative analysis of observed and synthetic CMDs. \cite{bertelli1992star} offered a novel perspective by examining the morphology and star counts of CMD in the Large Magellanic Cloud (LMC). \cite{gallart1996local} extended the approach by including a wider range of parameters, allowing a more detailed characterization of star positions, sizes, and population densities in CMDs. \cite{tolstoy1996interpretation} introduced maximum likelihood methods for synthetic CMD fitting, although it still required manual intervention. 

The latter developed isochrone fitting or synthetic CMD fitting approach provides the most accurate and quantitative way to derive SFHs \citep{dolphin2002numerical, aparicio2009iac, cignoni2010star, cole2014delayed, garling2024measuring}. This method generates synthetic stellar population CMDs, depending on the theoretical isochrones libraries with various age and metallicity distributions. The observed and synthetic CMDs are partitioned into grids, allowing linear combinations and $\chi^2$ minimization \citep{dolphin1997new, aparicio1997star}. Currently, three prominent methods are commonly used in the field of CMD-based SFH measurement: \textsc{match} \citep{dolphin2002numerical, weisz2014star1, skillman2017islands}; Cole's approach \citep[hereafter \textsc{Cole07}, ][]{cole2007leo, lianou2013star}; and IAC-star/IAC-pop \citep[hereafter \textsc{IAC}, ][]{aparicio2009iac, monelli2010acs6}. 

\textsc{match} \citep{dolphin2002numerical} aims to determine the SFH, distance, extinction, and chemical enrichment evolution that best reproduces the observed CMD. The advantage of \textsc{match} is that it extracts multiple physical parameters in CMDs together when the observational data are deep enough. However, when the observations are shallow, the high freedom of fitting parameters always leads to large uncertainties. \textsc{Cole07} \citep{cole2007leo, cole2014delayed} reduces the fitting uncertainty of the SFH by fixing parameters (distance and extinction) and scaling relations (age-metallicity relations) from other multi-wavelength observations. This method uses a special binary star recipe in which binaries are set to two types, "wide" and "close" binaries, with distinct initial mass function (IMF) distributions. \textsc{IAC} \citep{aparicio2009iac} draws inspiration from the earlier work of \cite{dolphin1997new} and \cite{aparicio1997star}. Its advantage is that it divides the CMDs into a combination of uniform and a la carte grids, allowing manual intervention depending on the well-known stellar evolution phases. These three approaches use Markov Chain Monte Carlo, Downhill Simplex, and Genetic Algorithm to achieve convergence, which are highly sensitive to the initial guess. To achieve faster convergence, the three methods adopt strong constraints, such as the initial assumptions of the SFH and the chemical enrichment law or fixed age-metallicity relations. \textsc{IAC} and the upcoming StarFormationHistories.jl \citep{garling2024measuring} are public, while other methods are not available to external users. 

It is noteworthy that most existing codes typically report random uncertainties, which are easily measured through Monte Carlo. More significant systematic uncertainties due to the model assumptions, such as the theoretical isochrones, IMF, and binary fraction, are usually ignored. \cite{dolphin2012estimation} measured the uncertainty of theoretical isochrones in SFHs and shows that the systematic uncertainty of theoretical isochrones can be 2-4 times larger than the random uncertainties, at some ages. However, a quantitative discussion of other system uncertainties is still lacking.

The launch of new space telescopes, including JWST \citep{gardner2006james}, EUCLID \citep{scaramella2022euclid}, and the upcoming CSST \citep{zhan2011consideration} and Roman \citep{mosby2020properties}, will facilitate a greater number of deep stellar-resolved photometries for nearby galaxies in the coming years. It is highly desirable to conduct comprehensive CMD studies using these datasets. There is a clear need for an open-source package with customizable input parameters, realistic uncertainty estimation, and high performance.

In this paper, we introduce an open-source tool, Python bAsed Numerical Color-magnitude-diagram Analysis pacKagE (\textsc{pancake}), designed to precisely measure the stellar population parameters in nearby galaxies. \textsc{pancake} is available on GitHub\footnote{https://github.com/Yunzheng168/PANCAKE} and Science BD \citep{scienceBDcode}. The algorithm of \textsc{pancake} draws inspiration from these three prominent methods, \textsc{match}, \textsc{Cole07}, and \textsc{IAC}. \textsc{pancake} includes three main tasks, i.e. template CMDs generation, CMD gridding, and CMD fitting (Section \ref{sec:CMDmethod}). In the gridding task, \textsc{pancake} provides three gridding methods (uniform binning, quadtree binning, and Voronoi binning) to meet the requirements of different proposals. In the fitting task, we apply a model-free fitting using \textsc{PyTorch} with gradient descent and backpropagation \citep{paszke2019pytorch}. The fitting process can converge within minutes on an 8-core CPU, with no initial constraints on ages and metallicities.

%%%%%%%%%%%%%
\begin{figure*}[ht!]
% \epsscale{0.9}
\plotone{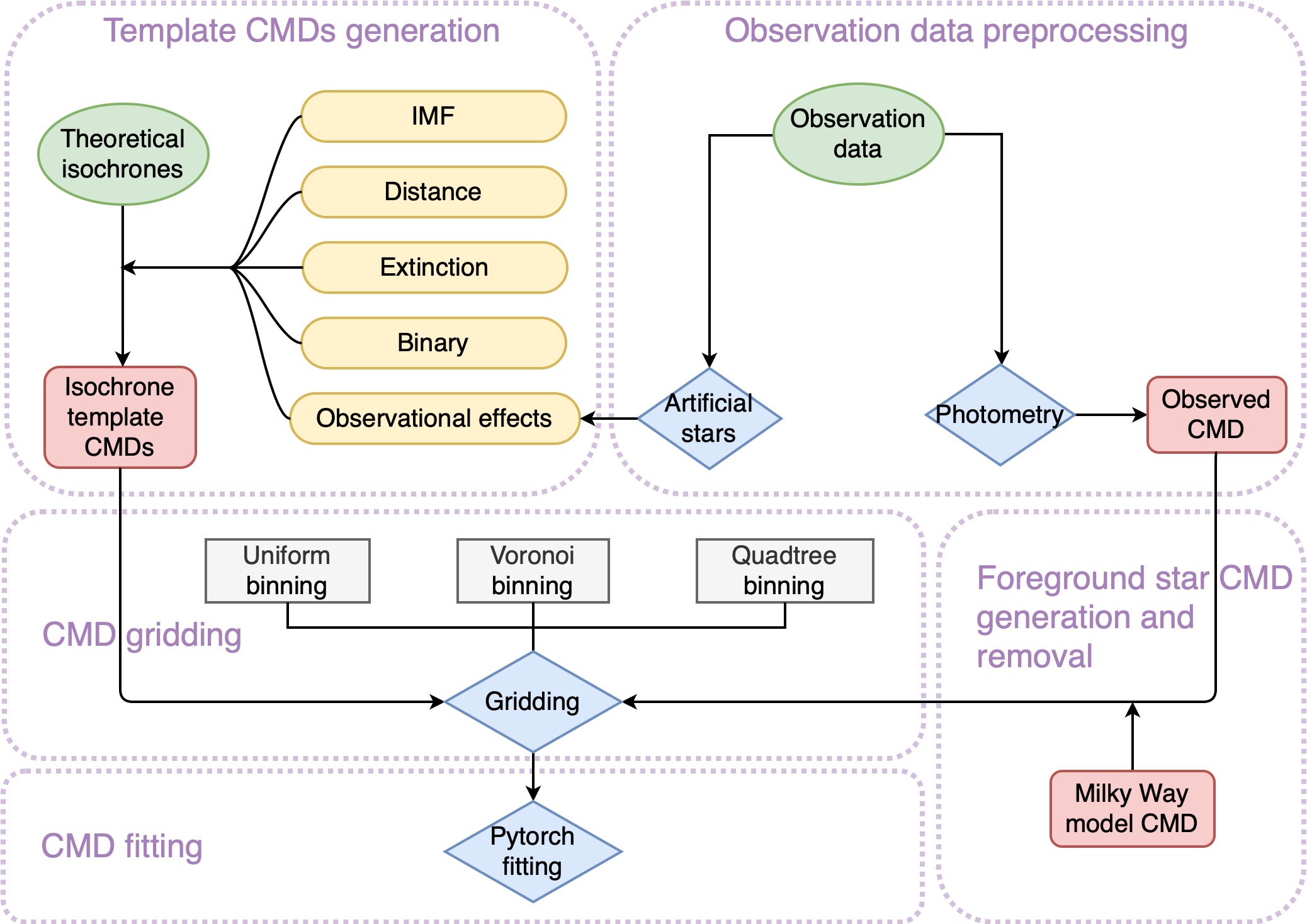}
\caption{The \textsc{pancake} workflow includes three main tasks (template CMDs generation, CMD gridding, and CMD fitting) and several assistant tasks (preprocessing, foreground star CMD generation and removing). A general description is given at the beginning of Section \ref{sec:CMDmethod}. A detailed introduction to each main task is given in the following subsections.}
\label{fig:workflow}
\end{figure*}
%%%%%%%%%%%%%

The validation of \textsc{pancake} is conducted through a series of mock CMDs, which allow quantifying the uncertainties introduced by the variated gridding methods, IMFs, binary fractions, star counts, and isochrones (Section \ref{sec:mock}). 
% We skip the uncertainty of theoretical isochrones as it is already well discussed in \cite{dolphin2012estimation}. 
We analyze the uncertainties from typical 1D cumulative SFHs, and innovative 2D age-metallicity maps which include distribution information in both time and metallicity.

\textsc{pancake} is further validated with observational data. First, we apply \textsc{pancake} to a star cluster in Section \ref{sec:starcluster}. Then, we analyze the dwarf galaxies from the \cite{weisz2014star1} dataset, allowing for a direct comparison with the widely used \textsc{match} method (Section \ref{sec:weisz}). Finally, we compare the star formation rate (SFR) from \textsc{pancake} and that from far-ultraviolet (FUV) photometry, as presented in Section \ref{sec:FUV}. Limitations of \textsc{pancake} are discussed in Section \ref{sec:limitation}. \textsc{pancake} is a powerful open-source tool for analyzing the stellar populations in nearby galaxies or star clusters. All of the input data, intermediate steps, and output results of this work are available in the Science BD (Appendix \ref{appendix_data}). 

%%%%%%%%%%%%%%%%%%%%%
\section{Method} \label{sec:CMDmethod}
\textsc{pancake} employs linear modeling to fit observed CMD data using a set of isochrone template CMDs, accounting for the influence of foreground stars in the Milky Way. The workflow is shown in Figure \ref{fig:workflow}. \textsc{pancake} includes three main tasks (template CMDs generation, CMD gridding, and CMD fitting) and several assistant tasks (preprocessing, foreground star CMD generation and removing). A detailed introduction of each main task is presented in the subsequent subsections. We clarify that the term `star counts' in the following text refers to the total number of data points in the CMD. The following is a general outline of the workflow:

\begin{itemize}
\setlength{\itemsep}{-1pt}
    \item {\bf Preprocessing:} Before running \textsc{pancake}, the observational data and theoretical isochrones (green boxes in Figure \ref{fig:workflow}) should be transformed into the \textsc{pancake}-specific format (see details in Appendix \ref{appendix_format}). We allow the users to freely change the observational dataset and the theoretical isochrone set. Users need to do their own pre-processing to convert to the required data format. \textsc{pancake} provides a preprocessing code for LOGPHOT data and PARSEC isochrone set. For observational data, \textsc{pancake} provides the preprocessing of LOGPHOT \citep[the Local Group Stellar Photometry Archive;][]{holtzman2006local}. Theoretical isochrones are derived from the widely used stellar evolutional model. \textsc{pancake} provides the preprocessing of the stellar evolutional model, PAdova and TRieste Stellar Evolution Code \citep[PARSEC;][]{bressan2012parsec}.
    \item {\bf Template CMDs generation:} Isochrone template CMDs are created by combining theoretical isochrones with input parameters, including IMF, distance, extinction, binary, and observational effects. Yellow boxes in Figure \ref{fig:workflow} show these input parameters. \textsc{pancake} provides three IMFs for user selection, Kroupa \citep{kroupa2001variation} and Chabrier individual and system \citep{chabrier2003galactic}. Observational effects (crowding, blending, observational flux errors, etc.) could be corrected by the completeness and uncertainty curves detected by the artificial star test. The output is a set of isochrone template CMDs. The specific number of template CMDs depends on the age and metallicity intervals of the theoretical stellar evolutional model. 
    \item {\bf Foreground star CMD generation and removal:} The foreground star CMD is generated using the Milky Way stellar structure model \citep{de2010mapping}. These foreground stars are then removed from the observed CMD. This task is optional.
    \item {\bf CMD gridding:} \textsc{pancake} grids the 2D CMD distributions into 1D star-count arrays, transforming a complex 2D fitting problem into a simple 1D linear fitting. Three gridding methods are available, uniform binning, quadtree binning, and Voronoi binning. The former one grids the CMD uniformly, while the latter two grid the CMD according to its stellar density.
    \item {\bf CMD fitting:} \textsc{pancake} applies a model-free linear fitting using \textsc{PyTorch} with gradient descent and backpropagation \citep{paszke2019pytorch}. The fitting outputs are the number of stars formed in each template CMD at a specific age and metallicity. For further scientific analysis, the star counts need to be transformed into stellar mass or SFR, which is based on each theoretical isochrone and IMF distribution. The transformation matrix is a byproduct of the template CMDs generation task.
\end{itemize}

It takes about 30 minutes to run \textsc{pancake} on an 8-core CPU for a typical galaxy with 50,000 star counts. The preparation files are the IMFs (2.4 GB), theoretical isochrones (700 MB), and the Milky Way stellar structure model (7.4 GB). The size of the \textsc{pancake} intermediate steps and output files is approximately 600 MB. 

%###########
\subsection{Template CMDs generation} \label{sec:CMDmethod:template}
Each template CMD consists of $10^4$ data points. Firstly, the template CMD generation task starts by randomly sampling the $10^4$ masses according to the IMF. There are three optional IMFs, Kroupa \citep{kroupa2001variation} and Chabrier individual and system \citep{chabrier2003galactic}. We then obtain the absolute magnitudes in two bands by linear interpolation of the mass-luminosity relation provided by the theoretical isochrones. The absolute magnitudes are convolved with the target distance and extinction, which are two input parameters. In this paper, we use the PARSEC \citep{bressan2012parsec}, which is widely used in the community and is under active development. 

We allow the prospective users to freely switch the theoretical isochrone, which should be transformed into the \textsc{pancake}-specific format (see details in Appendix \ref{appendix_format}).

Several stellar evolution models are available in the literature, such as PARSEC \citep{bressan2012parsec}, Dartmouth \citep{dotter2008dartmouth}, BaSTI \citep{pietrinferni2004large}. These models differ in many aspects, such as their age and metallicity ranges, adopted solar chemical compositions, opacities, and treatments of convective mixing. A detailed comparison of these stellar evolution tracks can be found in \cite{bressan2012parsec}. Below, we summarize their parameter spaces for user reference.

Dartmouth provides isochrones of stellar masses between 0.1 - 4 $M_\odot$ and stellar ages between 0.25 - 15 Gyr. This model spans a range of [Fe/H] = -2.5 to +0.5 dex and a range of helium mass fractions Y = 0.245 to 0.40. BaSTI extends to a broader mass range (0.5 to 10 $M_\odot$) and younger ages (30 Myr - 15 Gyr). Its metallicity range is [Fe/H] = -2.27 to +0.40 dex, with initial Helium mass fraction Y = 0.245 to 0.303. As the latest model, PARSEC has the widest mass and metallicity ranges. The tracks are computed for a scaled-solar composition and following the Y = 0.2485 + 1.78$Z$, with $Z$ ranging from 0.02 to 0.03. PARSEC also incorporates a wide array of photometric systems for observational comparisons. We analyze the systematic errors associated with different isochrones in Section \ref{sec:Mock_iso}.

%%%%%%%%%%%%%
\begin{figure*}[ht!] 
% \epsscale{0.9}
\plotone{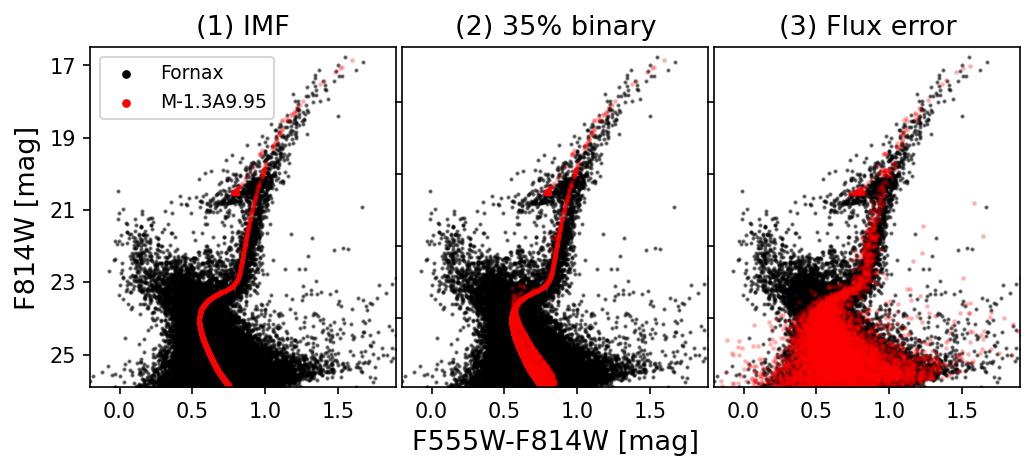}
\caption{Schematic diagram of an isochrone template CMD generation. The black points represent the CMD of the Fornax dwarf galaxy. The red points represent the isochrone template with ${\rm log}\ (t/ \rm yr) = 9.95$ ($t =$ 8.91 Gyr) and $\rm [M/H] = -1.3$. The three panels from left to right illustrate the steps of template CMD generation: (1) randomly sampling the $10^4$ masses according to the IMF, linearly interpolating the mass-luminosity relation provided by the theoretical isochrone, and convolving with the target distance and extinction; (2) randomly sampling 35\% binaries and adding companions; (3) incorporating observational flux errors and completeness.}
\label{fig:templatebuild}
\end{figure*}
%%%%%%%%%%%%%

Secondly, the binary stars are randomly selected with a free input parameter binary fraction $f$. The binary fraction $f$ determines that $f*10^4$ data points in CMD represent unresolved binary systems, while $(1-f)*10^4$ points remain single stars. These binary systems are randomly selected from the $10^4$ points of the template. The masses of $f*10^4$ companion stars are sampled from the main sequence mass range according to the same IMF. We linearly interpolate the companion stars' masses into the same isochrone and obtain the absolute magnitudes in two bands. The $f*10^4$ companion stars are then randomly paired with the original points in the template, and their magnitudes are added to the original stars' magnitudes in both bands. This causes the positions of the original points in the CMD to shift upward and to the right. We prefer to follow the $35\%$ binary star recipe in \textsc{match}, since the binary fraction ($65\%$) applied in \textsc{Cole07} is much higher than the typical value in the Milky Way. \cite{duchene2013stellar} shows that the multiplicity frequency increases with stellar mass, specifically, the binary fraction for solar-type stars and low-mass stars are around $41 - 50\%$ and $26\%$, respectively. The high binary fraction may lead to an overestimation of the masses of stellar populations. 

Finally, we incorporate observational flux errors and completeness derived from artificial stars in both bands, to ensure the templates CMDs have the same observational effects (crowding, blending, observational flux errors, etc.). Both observational flux errors and completeness derived from artificial stars are provided in the preprocessing, where they should be converted into the \textsc{pancake}-specific format (See details in Appendix \ref{appendix_format}).

Figure \ref{fig:templatebuild} illustrates the schematic diagram of constructing an isochrone template CMD. The background black points represent the CMD of the Fornax dwarf galaxy. The red points indicate an isochrone template CMD with ${\rm log}\ (t/ \rm yr) = 9.95$ ($t =$ 8.91 Gyr) and $\rm [M/H] = -1.3$. In this example, we use Kroupa IMF and 35\% binary fraction. 

%%%%%%%%%%%%%
\begin{figure*}[ht!]
% \epsscale{0.9}
\plotone{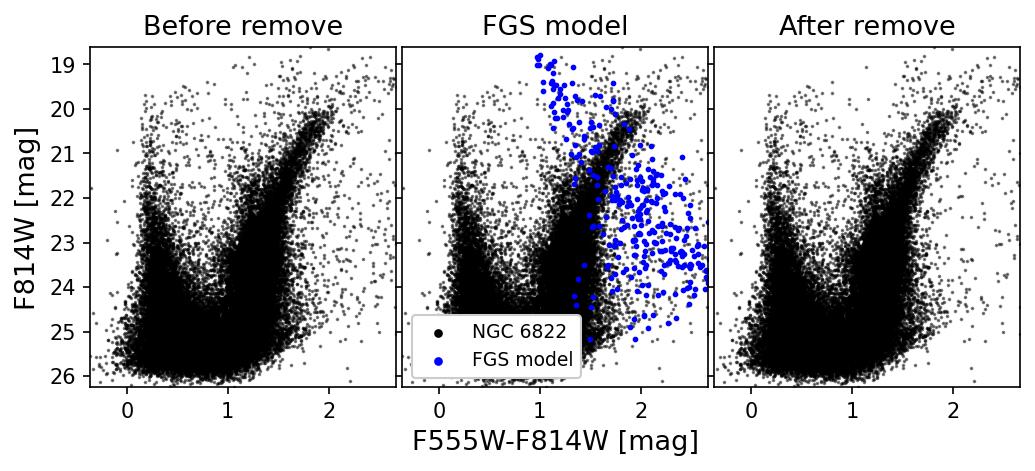}
\caption{Schematic diagram of the Milky Way foreground stars removal. The black dots in the first and second images show the original observed CMD of NGC 6822, which is a good example due to its large number of foreground stars. The blue points in the second panel represent the model CMD of foreground stars. To remove these foreground stars from the original observed CMD, we implement a removal criterion with $\delta_{\rm mag} < 1.5 \ \delta_{max}$, see text for details. The third panel shows the observed CMD after removing the foreground stars.}
\label{fig:FGS}
\end{figure*}
%%%%%%%%%%%%%

%%%%%%%%%%%%%%
\subsection{Foreground stars in the Milky Way} \label{sec:CMDmethod:foreground}
Considering the foreground stars, we draw inspiration from \textsc{match} and develop a Milky Way stellar population model based on the stellar structure \citep{de2010mapping}. In alignment with \cite{de2010mapping}, we adopt three stellar populations with different metallicities ([Fe/H] = -0.7, -1.3, and -2.2) and an age range of $ 10.1< {\rm log}\ (t/ \rm yr)<10.2$, corresponding to the thick-disk-like, inner-halo-like, and outer-halo-like components, respectively. The Milky Way stellar population model is established within a cylindrical coordinate system ($R,\theta, Z$). Notably, we exclude the thin-disk-like components and focus on regions with $Z > 1$ kpc, while we set $R < 40$ kpc considering the size of the Milky Way's halo. 

Figure \ref{fig:FGS} displays the schematic diagram of the Milky Way foreground stars removal. The black points are the observed CMD of NGC 6822, and the blue points are the foreground star CMD from the Milky Way model. For each foreground star in the CMD, we identify the nearest observed CMD star and calculate the distance in magnitude $\delta_{\rm mag}$ between them. Simultaneously, we compute the maximum error induced by observational flux errors $\delta_{\rm max} = \sqrt{\delta_{\rm color}^2+ \delta_{\rm magnitude}^2}$. If the distance in magnitude is less than 1.5 times the maximum error ($\delta_{\rm mag} < 1.5 \ \delta_{\rm max}$), we remove the nearest observed star. We assume a simple circular shape error region in CMD to remove the foreground stars. However, the errors of color and magnitude are not independent, which means that the accuracy error region should be elliptical. Fortunately, \textsc{pancake} can separate the remaining most likely foreground population (see Section \ref{sec:starcluster}) to compensate for the error introduced here.

%%%%%%%%%%%%%%%
\subsection{CMD gridding} \label{sec:gridding}
Before fitting the template CMDs to the observed one, we grid them into discrete bins. \textsc{pancake} provides three gridding methods, uniform binning, quadtree binning, and Voronoi binning (see Table \ref{tab:gridding}). 

The uniform binning utilizes the histogram2d function from the NumPy \citep{harris2020array} package, while the quadtree binning employs the qthist2d\footnote{https://github.com/jradavenport/qthist2d} package. Quadtree is a tree-like data structure where each internal node is associated with exactly four child nodes. Quadtree binning requires a threshold of star counts so that a rectangular bin will be subdivided into four child bins along the midpoints of color and magnitude axes. This hierarchical binning allows for adaptive grid refinement, ensuring that densely populated CMD regions receive more detailed binning, while sparser regions receive coarser binning. Since the star counts in the CMD follow a Poisson distribution, the error of the star counts (S) in each bin is $\sqrt{S}$. Finally, quadtree binning ensures that each bin contains similar values for star count and error, which benefits the subsequent fitting process. Voronoi binning has the same adaptive grid refinement but employs a different tessellation. The detailed algorithm of Voronoi binning is:

%%%%%%%%%%%%%
\begin{deluxetable*}{ccccc}[ht!]
% \vspace{-1cm}
\tablenum{1}
\tablecaption{Three CMD gridding methods \label{tab:gridding}}  
\tablewidth{0pt}
\tablehead{
\colhead{Methods} & \colhead{Input parameter} & \colhead{Explanation}}
\startdata
Uniform binning & Numbers of bins in color and magnitude axes & The number of bins is larger, the binning is denser. \\
Quadtree binning & Threshold of star counts & The threshold is smaller, the binning is denser.\\
Voronoi binning & Threshold of star counts & The threshold is smaller, the binning is denser.\\
\enddata
\tablecomments{
Using a galaxy with 11,000 star counts as an example, uniform binning with 100 bins in both color and magnitude axes, means the total grids are $\rm color \ bins \times magnitude \ bins = 100\times100 = 10,000$. While the quadtree and Voronoi binning with the threshold of 10 points means the total grids are roughly $\rm star count/threshold = 11,000/10 = 1,100$.
}
\end{deluxetable*}
%%%%%%%%%%%%%
%%%%%%%%%%%%%
\begin{figure*}[ht!]
% \epsscale{0.9}
\plotone{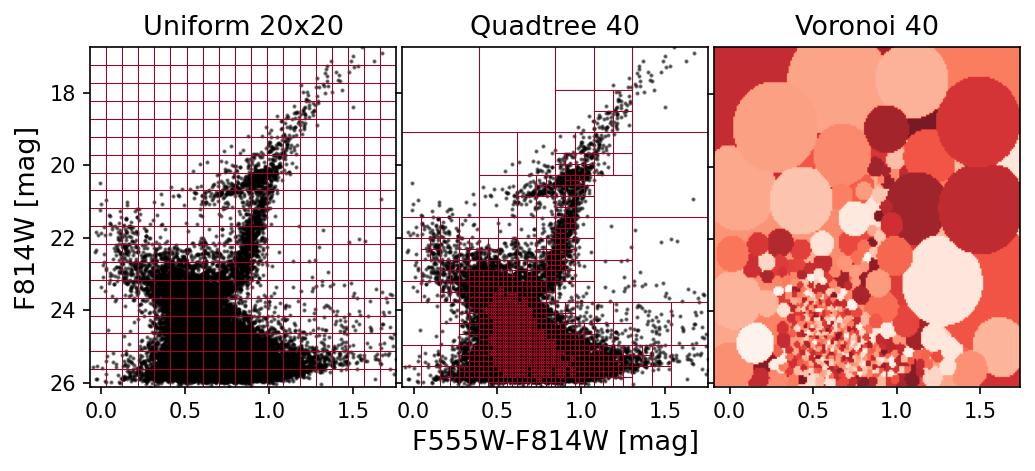}
\caption{Schematic diagram of three gridding methods, uniform binning, quadtree binning, and Voronoi binning. The background black points are the Fornax dwarf galaxy. Uniform binning has 20 bins in both color and magnitude axes, while quadtree binning and Voronoi binning have a threshold of 40 star counts.}
\label{fig:grid_ex}
\end{figure*}
%%%%%%%%%%%%%

\begin{itemize}
\setlength{\itemsep}{-1pt}
    \item[\bf a.] In preparation, Voronoi binning first uniformly grids the CMD into small pixel-bins.
    \item[\bf b.] The first bin starts with the densest pixel-bin, and the centroid and star count of that pixel-bin are stored.
    \item[\bf c.] A candidate pixel-bin is selected as the closest one to the current bin. The candidate is then added into the current bin, and the new centroid and star count are restored. 
    \item[\bf d.] A bin ends with criteria when either its star count or roundness exceeds the threshold. The roundness is calculated by $R = r_{\rm max}/r_{\rm eff}-1$, in which $r_{\rm max}$ is the maximum distance between the centroid and the edge of this bin, and $r_{\rm eff}$ is the radius of a circle with the same area as this bin. This criterion regulates the bin shape to avoid a bin crossing too many stellar population regions. 
    \item[\bf e.] If the criteria are met, the new bin begins from the nearest unbinned pixel-bin. 
\end{itemize}

Figure \ref{fig:grid_ex} demonstrates these three gridding methods using the Fornax dwarf galaxy as an example. In make the grids distinguishable in the figure, we chose relatively loose thresholds. Uniform binning has 20 bins in both color and magnitude, while quadtree binning and Voronoi binning have a threshold of 40 points. In real operations, smaller thresholds are required to achieve more accurate fitting results.

%%%%%%%%%%%%%%%
\subsection{CMD fitting}
After gridding the observed CMD into $N$ distinct bins, we count the points in each bin and yield a matrix $\bf D$ of dimensions $1\times N$. We employ the same binning to $I$ isochrone template CMDs and obtain a matrix $\bf M$ of dimensions $I \times N$. We apply a linear fitting to produce the weight matrix $\bf W$ for these isochrone template CMDs. The inputs are the matrices $\bf D$ and $\bf M$, while the output is the matrix $\bf W$. The fitting function is expressed as follows:

\begin{equation}
    {\bf D}_{1\times N} = {\bf W}_{1\times I} \ {\bf M}_{I\times N}.
\end{equation}

Considering the intricacy and speed of the fitting process, we utilize \textsc{PyTorch} with gradient descent and backpropagation \citep{paszke2019pytorch}. The loss function is selected as the Mean Squared Error, while the optimizer \textsc{Adam} \citep{kingma2014adam} is used with a learning rate of 0.001. Since the weights cannot be negative, we set all values that are less than 0 in the weight matrix $\bf W$ to 0 at the beginning of each epoch. To ensure convergence in the fitting process, the number of epochs is set to 50000, which is exceptionally large. Nevertheless, one fitting process can still be completed in a few minutes on an 8-core CPU, depending on the number of bins $N$.

%%%%%%%%%%%%%%%
\subsection{Star counts to mass ratio}
The fitting process provides the weight matrix $\bf W$ for these isochrone template CMDs. The number of stars formed in each isochrone ${\bf S}$ can be calculated by
\begin{equation}
{\bf S}_{1\times I} = \frac{1}{{\bf C}_{1\times I}} \cdot
  \sum_{n=1}^{N} {\bf M}_{in}  \cdot
 {\bf W}_{1\times I},
\end{equation}
in which the matrix $\bf C$ is the observational completeness of isochrones. The completeness of isochrones $\bf C$ describes the difference between the star counts in CMD and the actually formed stars, which is caused by the observational effects and binaries. 

The number of stars ${\bf S}$ can be transformed into star formation mass (SFM) or stellar mass (SM) through the IMF $f(m)$. SFM describes the initial star formation mass at a specific age and metallicity, while SM considers the mass loss during stellar evolution and describes the current stellar mass. We describe the SFH or SFR of galaxies using SFM and the current stellar mass of galaxies using SM.

First, for each isochrone $i$, the number of stars $S_i$ is in the mass range of this theoretical isochrone $m_1$-$m_2$, which is also bounded by the observed color and magnitude ranges. It needs to be extended to the mass range of IMF $M_1$-$M_2$, which is typically 0.08-150 $M_{\odot}$. 
\begin{equation}
\frac{S_i}{\int_{m_1}^{m_2}f(m) \ dm} = \frac{S_{i, \rm IMF}}{\int_{M_1}^{M_2}f(m) \ dm}.
\end{equation}

The SFM is the multiplication of the number of stars in the IMF mass range $S_{i, \rm IMF}$ and the average mass per star $\overline{M_{i, \rm IMF}}$, 
\begin{equation}
{\rm SFM_i} = {S_{i, \rm IMF}} \cdot \overline{M_{i, \rm IMF}},
\end{equation}
in which
\begin{equation}
\overline{M_{i, \rm IMF}} = \frac{\int_{M_1}^{M_2}m \ f(m) \ dm}{\int_{M_1}^{M_2}f(m) \ dm}.
\end{equation}

From equations (3-5), SFM-to-$\bf S$ ratio can be expressed by 
\begin{equation}
\frac{\rm SFM_i}{S_i} =  \frac{\int_{M_1}^{M_2}m \ f(m) \ dm}{\int_{m_1}^{m_2}f(m) \ dm}.
\end{equation}

When calculating the SM-to-$\bf S$, we consider the mass loss during the stellar evolution, and use the actual mass instead of the IMF mass $m$. The actual mass is a function of the IMF mass $A(m)$, which is provided in the theoretical isochrone.  
\begin{equation}
\frac{\rm SM_i}{S_i} = \frac{\int_{M_1}^{M_2}A(m)f(m) \ dm}{\int_{m_1}^{m_2}f(m) \ dm}.
\end{equation}

SFM-to-$\bf S$ and SM-to-$\bf S$ are calculated for each isochrone with dimensions $1\times I$. These two matrices and the completeness matrix $\bf C$ are the output parameters of the template CMDs generation task in Section \ref{sec:CMDmethod:template}. Here the value of $I = t \times z$ is the number of isochrones, depending on the age bins $t$ and metallicity bins $z$ set for the theoretical stellar evolutional model. The fitting output matrix $\bf S$ in the unit of number of stars or matrix $\bf SFM$/$\bf SM$ in the unit of stellar mass with dimensions $z \times t$ is hereafter called the age-metallicity map.

%%%%%%%%%%%%%###################
\section{Mock test} \label{sec:mock}
We use mock galaxies to evaluate the uncertainties of \textsc{pancake}. We construct 9 mocks based on parameters of Fornax dwarf galaxy (distance, extinction, observational effects, see Section \ref{sec:weisz:data}), as it has extremely deep photometry. The input age-metallicity maps are constructed from the SFH functions and age-metallicity relations. SFH functions include a continuum (constant, exponential, or delay) star formation and one or more Gaussian-distribution starbursts. The age-metallicity relation is constant or linear. 

The mock age-metallicity maps have 51 logarithmic time bins within the range of $ 6.6 \leq {\rm log}\ (t/ \rm yr)\leq 8.7$ with intervals of 0.1, and of $ 8.7 \leq {\rm log}\ (t/ \rm yr)\leq 10.15$ with intervals of 0.05, and have 24 logarithmic metallicity bins within the range of $ -2.2 \leq {\rm [M/H]} \leq 0.1$ with intervals of 0.1. 

The process of constructing mock CMD follows the template CMDs generation task (Section \ref{sec:CMDmethod:template}). Inputs are the model age-metallicity map ($\bf True$), IMF, and binary fraction. All the mock CMDs are built based on the Kroupa IMF \citep{kroupa2001variation} and a binary fraction of 35\%. We then apply \textsc{pancake} to the mock CMD and obtain an output age-metallicity map. We repeat these two processes (mock CMD generation and run \textsc{pancake}) and obtain a series of output age-metallicity maps. We use the mean value of the age-metallicity map as the final output ($\bf Output$) and use the standard deviation (STD) as the error ($\bf \sigma$). 

%%%%%%%%%%%%%
\begin{figure*}[ht!]
% \epsscale{0.8}
\plotone{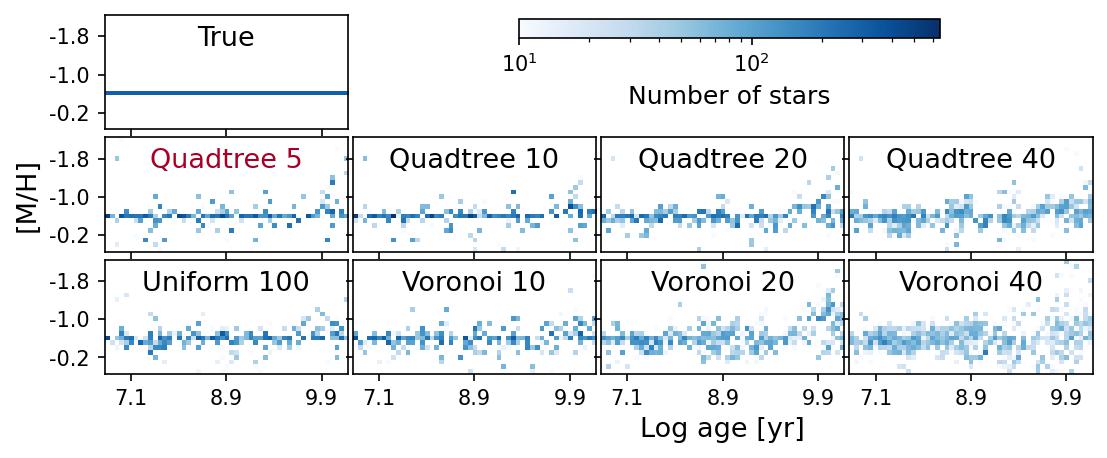}
\caption{The best-fit output age-metallicity maps for three gridding methods. The example mock CMD has constant SFH and constant metallicity. We use the Kroupa IMF and a binary fraction of 35\%. The uniform binning has 100 bins in both color and magnitude axes. The thresholds of quadtree and Voronoi binnings vary from 10 to 40. The normalized $\chi^2$s (in Figure 7) quantitatively show `Quadtree 5' as the best gridding recipe, here labeled in red.}
\label{fig:grid_com1}
\end{figure*}
%%%%%%%%%%%%%

The mock is constructed based on the SFH and age-metallicity relation model functions. If the CMD fitting method is also model-dependent, the different assumptions between the SFH model functions will lead to significant uncertainties during the mock test process. Fortunately, \textsc{pancake} applies model-free fitting to avoid this error. The best-fit output results in the following indicate that \textsc{pancake} could recover both the SFH and age-metallicity relation very well, despite using a model-free fitting.

Uncertainties can be classified into three categories. The most extensively debated uncertainty in the literature is random uncertainties (or fitting error), which can be estimated using the standard bootstrap Monte Carlo technique or Markov Chain Monte Carlo \citep{dolphin2013estimation}. The systematic uncertainties are attributable to the assumptions in \textsc{pancake} process, e.g. the template CMD generation (theoretical isochrones \citep{dolphin2012estimation}, IMF, binary fraction) or CMD gridding (methods, threshold). The final uncertainties are attributable to the different mock input age-metallicity maps, and the resulting different CMD distributions and densities (or star counts). 

To quantify the uncertainties, we describe the difference between the true mock input ($\bf True$) and \textsc{pancake} output ($\bf Output$) age-metallicity maps by the statistics value $\chi^2 = \sum \frac{ ({\bf True} - {\bf Output})^2 }{{\bf \sigma}^2} $. A smaller $\chi^2$ means the output result of \textsc{pancake} is closer to the actual true value of the mock. The random uncertainty of each mock is under consideration in $\bf \sigma$. The uncertainty for star counts in CMD is discussed in Section \ref{sec:mockNUM}. For the other uncertainties, the 9 mocks all have $\sim$11,000 points in CMD. To make 9 mocks comparable, we normalize the age-metallicity map by $10^{3.5} /\sum \bf True$. The constant $10^{3.5}$ is used to make it easier to display the relative values between $\chi^2$s.
\[
{\rm normalized} \ \chi^2 =\rm \frac{10^7}{(\sum {\bf True})^2} \sum \frac{ ({\bf True} - {\bf Output})^2 }{{\bf \sigma}^2} 
\]

\subsection{Gridding effect} \label{sec:gridtest}
Firstly, we evaluate the influence of three gridding methods, uniform binning, quadtree binning, and Voronoi binning. The uniform binning has 100, 50, and 20 bins in both color and magnitude axes. The thresholds of quadtree and Voronoi binnings refer to the maximum number of points in each bin and vary from 10 to 40. 

\begin{figure}[h!]
% \vspace{-5pt}
% \epsscale{0.9}
\plotone{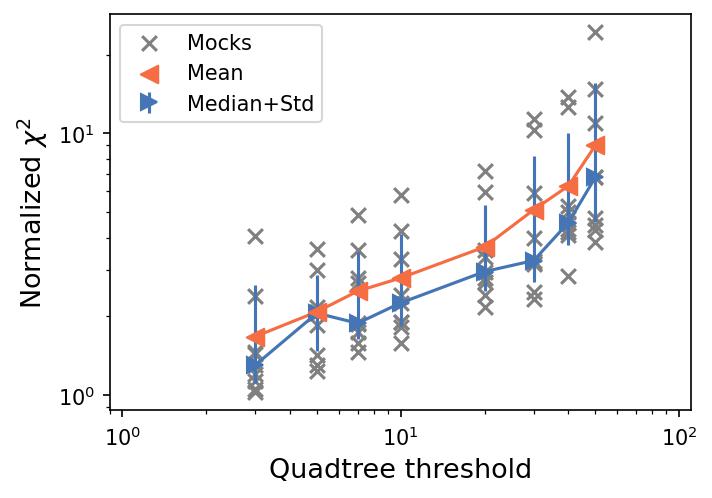}
\caption{The normalized $\chi^2$s for various thresholds of quadtree binning. The mocks are in gray. The mean is in orange, while the median and STD values are in blue.}
\label{fig:grid_com_quad}
\end{figure}

Using a mock CMD with constant SFH and constant metallicity as an example, the true and output age-metallicity maps are shown in Figure \ref{fig:grid_com1}. The output distribution of the quadtree binning with threshold 5 has the smallest scatter compared to the true distribution. Although the advantage of `Quadtree 5' over `Quadtree 10' is not obvious in the age-metallicity relation shown in Figure \ref{fig:grid_com1}. Figure \ref{fig:grid_com_quad} shows that the normalized $\chi^2$s increases monotonically with increasing threshold of the quadtree binning. Therefore, we recommend `Quadtree 5' as the best binning recipe. Voronoi binning produces a much larger scatter in metallicity. 

%%%%%%%%%%%%%
\begin{figure}[h!]
% \vspace{-5pt}
% \epsscale{0.9}
\plotone{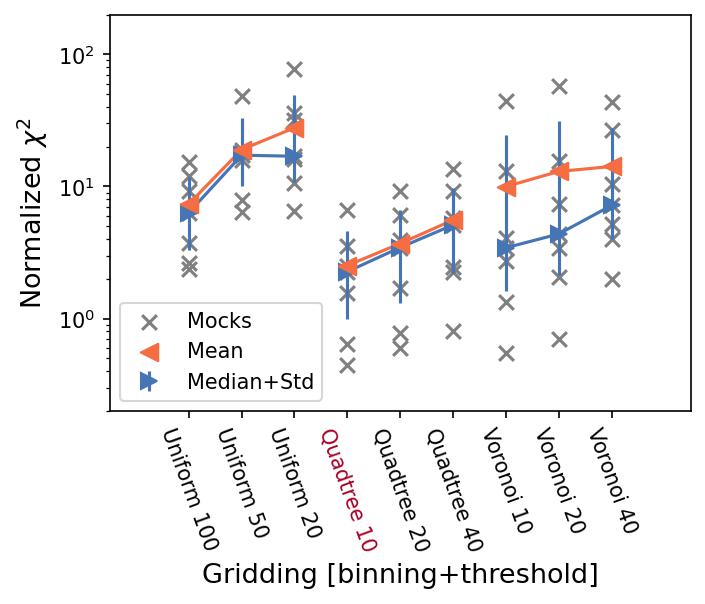}
\caption{The normalized $\chi^2$s for various gridding methods. The mocks are in gray. The mean is in orange, while the median and STD values are in blue. The relative best gridding recipe (quadtree with threshold 10) is labeled in red.}
\label{fig:grid_com2}
\end{figure}
%%%%%%%%%%%%%

Figure \ref{fig:grid_com2} shows the normalized $\chi^2$s for various gridding methods. The gray points are mocks. The orange points are the means, while the blue points are the medians and STD errors. For each binning method, the normalized $\chi^2$s become larger as the threshold becomes looser. 

We compare the most refined threshold of three binnings. Uniform binning with a threshold of 100 corresponds to 10,000 bins. Since each mock has nearly 11,000 points, quadtree and Voronoi binnings with thresholds of 10 correspond to $\sim$1,100 bins. More bins consume more fitting time. The quadtree and Voronoi binnings are more efficient than the uniform binning, as they have fewer bins and smaller normalized $\chi^2$s.

The uncertainty for CMD distribution has a strong correlation with the gridding method, which can be indicated in the STD error and the offset between mean and median. The Voronoi have large STD errors and large differences between mean and median, which means that Voronoi binning is heavily influenced by the uncertainty for CMD distribution. 

In summary, the quadtree binning is the relatively best gridding recipe in Figures \ref{fig:grid_com1} and \ref{fig:grid_com2}.

%%%%%%%%%%%%%
%%%%%%%%%%%%%

Then, we further test the quadtree binning with different thresholds in Figure \ref{fig:grid_com_quad}. Both the STD errors and the difference between the values of means and medians are stable, which implies that the influence of uncertainty for CMD distribution is small in quadtree binning. The normalized $\chi^2$ changes less when the threshold is below 10. Considering the normalized $\chi^2$s and the computational time consumption, we use the quadtree binning with the threshold 5 in the following sections. 

%%%%%%%%%%%%%
\begin{figure*}[ht!]
% \epsscale{0.95}
\plotone{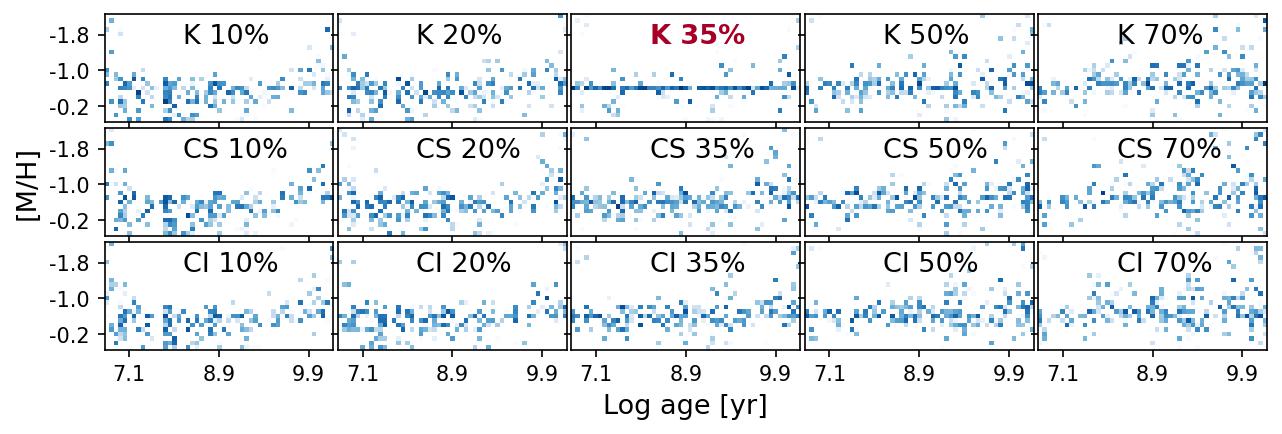}
\caption{The best-fit output age-metallicity maps for various IMFs and binary fractions. The example mock CMD has constant SFH and constant metallicity, and is built based on the Kroupa IMF and a binary fraction of 35\%. K, CI, and CS denote the Kroupa, Chabrier individual, and Chabrier system IMFs, respectively. The binary fractions range from 10\% to 70\%. The true age-metallicity map of example mock is shown in the upper-left panel of Figure \ref{fig:grid_com1}. When the IMF and binary fraction are the same as the input (Kroupa IMF and 35\% binary fraction; labeled in red), the fitting result is best.}
\label{fig:IMF_com1}
\end{figure*}
%%%%%%%%%%%%%

%%%%%%%%%%%%%
\begin{figure*}[ht!]
% \epsscale{0.95}
\plotone{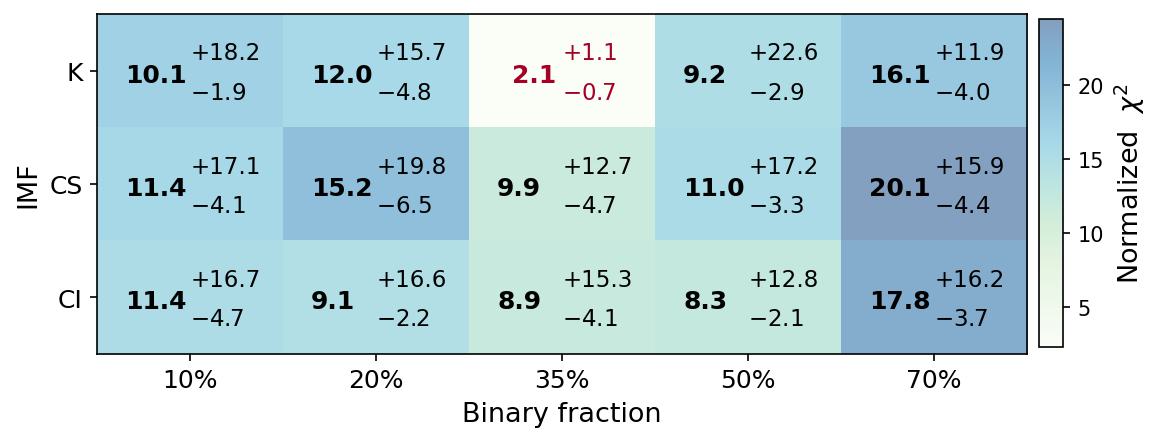}
\caption{The normalized $\chi^2$s for various IMFs and binary fractions. The mean values of normalized $\chi^2$s are presented in the color bar and labeled with errors. When the IMF and binary fraction are the same as the input (Kroupa IMF and 35\% binary fraction; labeled in red), the fitting result is best.}
\label{fig:IMF_com2}
\end{figure*}
%%%%%%%%%%%%%

\subsection{IMF and binary effect} \label{sec:mockIMF}
We discuss the influence of IMF which is based on the Kroupa IMF \citep[K;][]{kroupa2001variation} and Chabrier IMF individual (CI) and system \citep[CS;][]{chabrier2003galactic}. We test the binary fraction range from 10\%-70\%, which is the typical range for star clusters \citep{duchene2013stellar}. We also present the mock CMD with constant SFH and constant metallicity as an example to show the output age-metallicity maps with various IMFs and binary fractions in Figure \ref{fig:IMF_com1}, and the true age-metallicity map of the example mock is shown in the upper-left panel of Figure \ref{fig:grid_com1}. It is clear that when the IMF and binary fraction are the same as the input (Kroupa IMF and binary fraction 35\%), the fitting result is best. When the binary fraction is smaller than the true mock value, the young stellar population will show a larger scatter in metallicity. If the binary fraction is larger than the mock, the old stellar population will show a larger scatter in metallicity.

We calculate the normalized $\chi^2$s of 9 mocks and present their median and STD errors in Figure \ref{fig:IMF_com2}. Figure \ref{fig:IMF_com2} shows that on average the offset value of binary fractions is larger, and the normalized $\chi^2$s is larger.

%%%%%%%%%%%%%
\begin{figure}[h!]
% \vspace{10pt}
% \epsscale{0.9}
\plotone{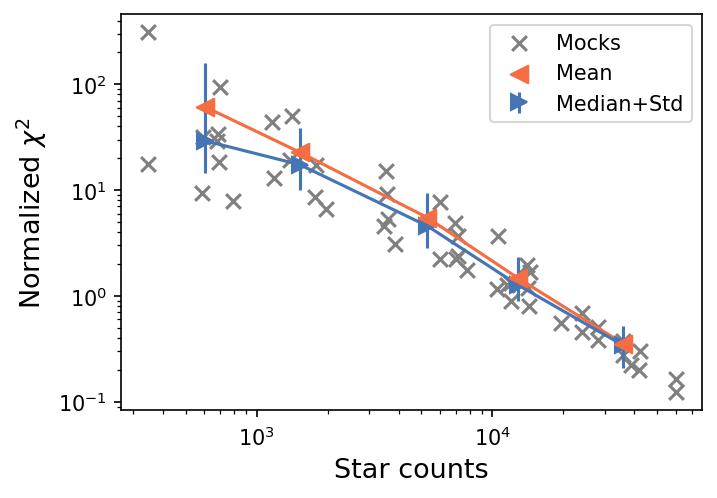}
\caption{The normalized $\chi^2$s for various star counts of mock CMDs. The mocks are in gray. The mean is in orange, while the median and STD values are in blue.}
\label{fig:stellarnum}
\end{figure}
%%%%%%%%%%%%%

\subsection{Observation star counts effect} \label{sec:mockNUM}
There is no doubt that the larger the amount of data, the more accurate the fitting will be. As the model star counts decrease, the uncertainty of the fit increases. We vary the model star counts by changing the intensity of the 9 mocks. This means that the true age-metallicity maps have the same overall distribution shape but different densities, and get rid of the uncertainty for CMD distributions. The normalized $\chi^2$s decrease as the star counts increase, as shown in Figure \ref{fig:stellarnum}. In the former tests, the 9 mocks all have $\sim$11000 points in CMD, corresponding to the median normalized $\chi^2 = 2.06$.

%%%%%%%%%%%%%%
\begin{figure*}[ht!]
% \epsscale{0.9}
\plotone{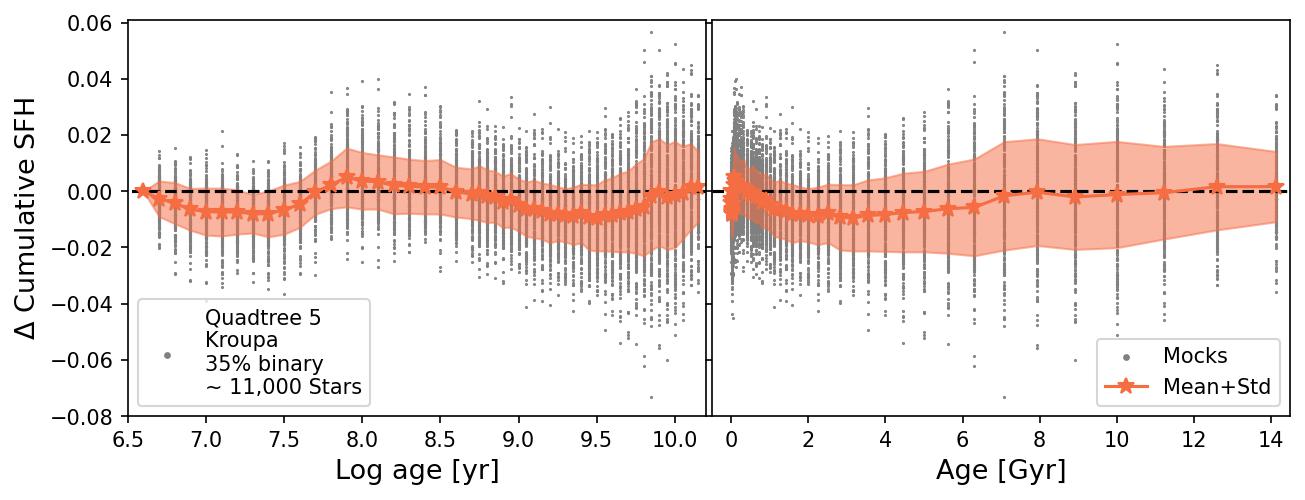}
\caption{The $\Delta$ cumulative SFH in mocks with $\sim$11,000 points, as the \textsc{pancake} input IMF and binary fraction are the same as mocks, and using quadtree binning with threshold 5 suggested in Section \ref{sec:gridtest}. The left panel shows the mocks with a logarithmic age x-axis in years, and the right panel shows the mocks with an age x-axis in Gyr.}
\label{fig:mock_csfh_ran}
\end{figure*}
%%%%%%%%%%%%%%

%%%%%%%%%%%%%%
\begin{figure*}[ht!]
% \epsscale{0.9}
\plotone{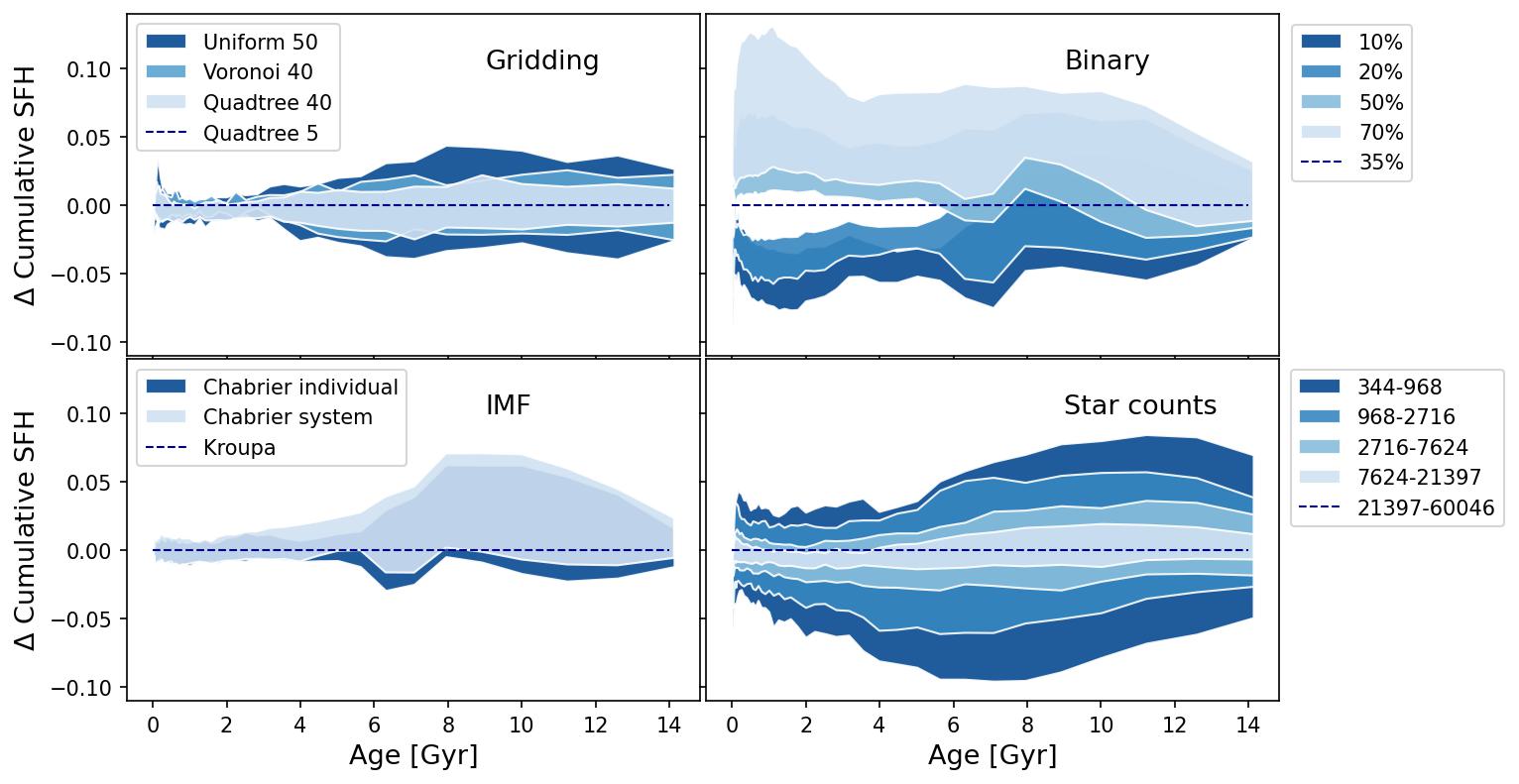}
\caption{Systematic uncertainties for gridding methods, binary fractions, IMFs, and the uncertainty for star counts (age-metallicity map densities). The uncertainties in the figure exclude the basic uncertainties (see text), therefore only the uncertainties themselves are included and quantified.}
\label{fig:mock_csfh_sys}
\end{figure*}
%%%%%%%%%%%%%%

\subsection{Uncertainty for cumulative SFH} \label{sec:mock_csfh}
In order to be comparable with the literature and to give a hint for the following cumulative SFH comparison results, we replace the 2D $\chi^2$ errors by the 1D $\Delta$ cumulative SFH errors in this subsection. The $\bf True$ and $\bf Output$ age-metallicity maps are first converted to cumulative SFH. The $\Delta$ cumulative SFH (Output - True) is calculated in each age bin. 

Figure \ref{fig:mock_csfh_ran} shows the $\Delta$ cumulative SFH in mocks with $\sim$11,000 points, where the \textsc{pancake} input IMF and binary fraction are the same as mocks, and using quadtree binning with threshold 5 suggested in Section \ref{sec:gridtest}. The mean offset from zero is the average difference from the true input, which represents the quantified uncertainties, while the STD errors represent the combination between the random uncertainty and the uncertainty for CMD distributions. These uncertainties in Figure \ref{fig:mock_csfh_ran} are hereafter referred to as basic uncertainties. The following analysis of the systematic uncertainties excludes the basic uncertainties, therefore only the systematic uncertainties themselves are included and quantified. Figure \ref{fig:mock_csfh_sys} shows the systematic uncertainties for gridding methods, binary fractions, IMFs, and the uncertainty for star counts (age-metallicity map densities). For systematic uncertainties, the means are given as $\rm mean = mean_0 - mean_{basic}$, while the STD errors are given as $\rm STD = \sqrt{STD_0^2-STD_{basic}^2}$. In the analysis of uncertainty for star counts, mocks are separated into 6 classes depending on their star counts. The mocks with the largest star counts (21,397 - 60,046 star counts) are expected to be the most accurate. Therefore, the basic uncertainties are calculated from the mocks with the largest star counts instead of the basic uncertainties in Figure \ref{fig:mock_csfh_ran} which has $\sim$11,000 star counts.

The upper left panel in Figure \ref{fig:mock_csfh_sys} compares three gridding methods, uniform binning with threshold 50, Voronoi binning with threshold 40, quadtree binning with threshold 40, and quadtree binning with threshold 5. The quadtree binning has the smallest scatter. For the mocks with $\sim$11,000 points, the relative numbers of bins are 2500, 275, and 275, respectively. The systematic uncertainty for gridding methods is the smallest of three systematic uncertainties in Figure \ref{fig:mock_csfh_sys}, and shows no preferential influence of stellar population age. 

The upper right panel in Figure \ref{fig:mock_csfh_sys} shows the systematic uncertainty for binary fractions. Since the input binary fraction is 35\% (black dashed line), the binary fraction largely influences the stellar population between 0-8 Gyr. If the binary fraction is smaller than true, the cumulative SFH is less than the expected value, typically between 0-8 Gyr. This means that the result presents too many young and middle-aged populations. If the binary fraction is larger than true, the opposite result occurs. 

The lower left panel in Figure \ref{fig:mock_csfh_sys} presents the systematic uncertainty for IMFs. The Chabrier IMFs increase the cumulative SFH in 7-12 Gyr populations, as the mock input IMF is Kroupa. 

The lower right panel in Figure \ref{fig:mock_csfh_sys} shows the uncertainty for star counts. As expected, the uncertainty increases with decreasing star counts. For mocks with the smallest star counts, the downward shift from zero is evident around 4-7 Gyr. Binary and star counts are two more significant effects for $\Delta$ cumulative SFH.

%%%%%%%%%%
\begin{figure*}[h!]
% \epsscale{0.9}
\plotone{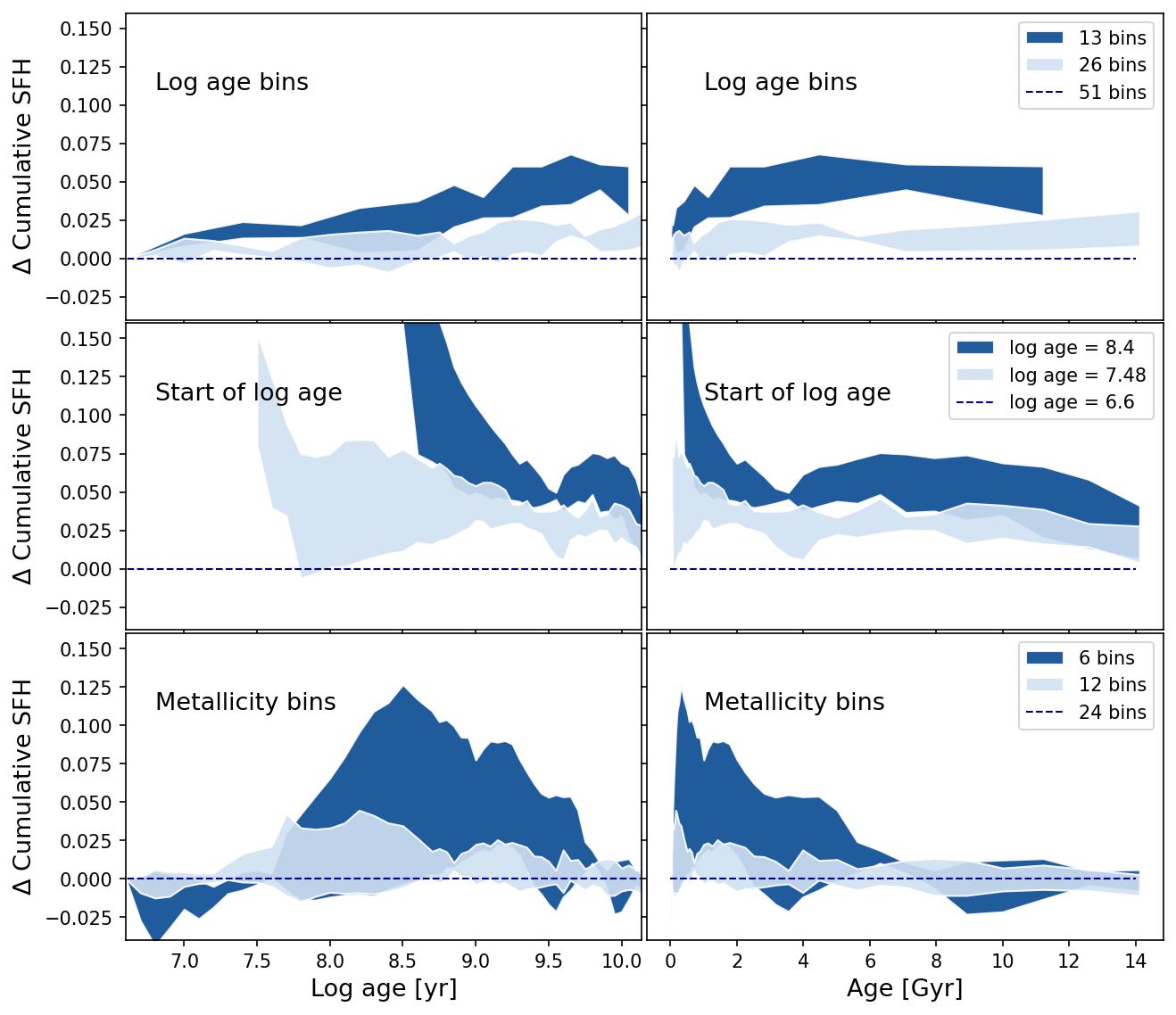}
\caption{Systematic uncertainties for varying logarithmic age bins, logarithmic age ranges, and metallicity bins using the PARSEC isochrone. The left panel has logarithmic age on the x-axis, while the right panel has age in Gyr. The top panel shows the results for different (13, 26, and 51) logarithmic age bins, the middle panel shows the effect of different initial values (6.6, 7.48, and 8.4) for the logarithmic age bins, and the bottom panel shows the results for different (6, 12, and 24) metallicity bins.}
\label{fig:isochrone_parsec}
\end{figure*}
%%%%%%%%%%
%%%%%%%%%%
\begin{figure*}[h!]
% \epsscale{0.9}
\plotone{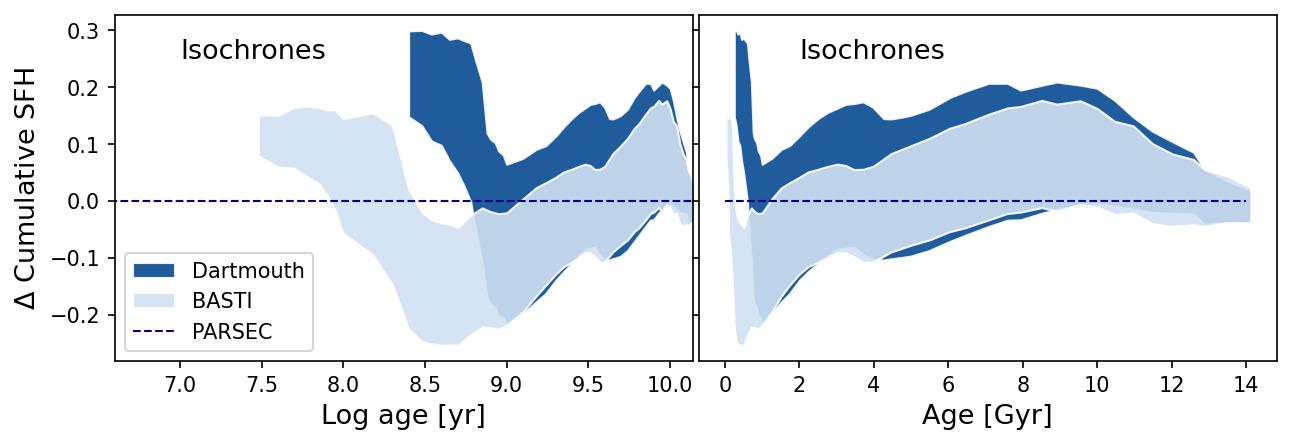}
\caption{Systematic uncertainties for various isochrones, Dartmouth, BaSTI, and PARSEC. Different isochrones lead to the largest systematic uncertainties in $\Delta$ cumulative SFH, ranging between -0.2 and 0.3.}
\label{fig:isochrones}
\end{figure*}
%%%%%%%%%%
%%%%%%%%%%%%%%%%%%%%%%%%%%
\subsection{Isochrones effect}\label{sec:Mock_iso}

In this subsection, we tested the effect of different isochrones. The detailed introduction of different isochrones can be found in Section \ref{sec:CMDmethod:template}. Because different isochrones provide different age and metallicity ranges. It's not feasible to describe the errors using the normalized $\chi^2$ parameter defined at the beginning of Section \ref{sec:mock}. Therefore, we continued to use the $\Delta$ cumulative SFH defined in Section \ref{sec:mock_csfh} to analyze the systematic errors of the isochrones.

We established 9 mocks with the theoretical isochrone, PARSEC, which has 51 age bins and 24 metallicity bins (see details in Section \ref{sec:mock}). Each Mock has $\sim$11,000 points, a Kroupa IMF, and 35\% binaries. To study the uncertainty of isochrones, we only changed the properties of the theoretical isochrones during fitting, and kept the other input parameters fixed (the Kroupa IMF, 35\% binary fraction, and quadtree binning with a threshold of 5). Similarly, to focus solely on the systematic errors introduced by the isochrones, we defined basic uncertainties that include random uncertainty and the uncertainty for CMD distributions. The basic uncertainty was determined by multiple fitting with the same input isochrone parameters, PARSEC (51 age bins × 24 metallicity bins), used in the mock-up.

Since the available logarithmic age bins, logarithmic age ranges, and metallicity bins vary across different isochrone models, we first tested these three effects using the PARSEC isochrone before comparing different isochrone sets. The PARSEC isochrone was initially set up with 51 age bins and 24 metallicity bins. When testing reduced bin numbers along one parameter axis (age/metallicity), we kept the other axis (metallicity/age) unchanged. Additionally, we change the starting logarithmic age from 6.6 to 7.48 or 8.4 yr, matching the start logarithmic age of the BaSTI (initial log age = 7.48 yr) and Dartmouth (initial log age = 8.4 yr) isochrones.

The results are shown in Figure \ref{fig:isochrone_parsec}. The x-axis of the left panel is logarithmic age, that of the right panel is linear age, in units of Gyr. The top two figures show the results of different (13, 26, and 51) logarithmic age bins. Reducing the number of logarithmic age bins results in a consistently higher $\Delta$ cumulative SFH across all age ranges. The deviation of $\Delta$ cumulative SFH increases with age, and the slope of the age-$\Delta$ cumulative SFH relationship becomes steeper as the number of logarithmic age bins decreases. When the number of logarithmic age bins is reduced to 26, the average $\Delta$ cumulative SFH is 0.01. When the number of logarithmic age bins is reduced to 26, the average $\Delta$ cumulative SFH is 0.03. The maximum systematic error caused by 13 logarithmic age bins does not exceed 0.075, which is smaller than the systematic errors caused by binary fractions and star counts.

The middle two figures show the effect of different starting values for the logarithmic age bins on the fitting results. It can be seen that changing the start of the logarithmic age from 6.6 to 7.48 or 8.4 yr leads to an increase in the $\Delta$ cumulative SFH over all age ranges. Moreover, this upward trend decreases with increasing age. When the start of logarithmic age bins is 7.48 yr, the average $\Delta$ cumulative SFH is 0.04. When the start of logarithmic age bins is 8.4 yr, the average $\Delta$ cumulative SFH is 0.07.

The bottom two figures show the results of different (6, 12, and 24) metallicity bins. When the number of metallicity bins is reduced from 24 to 12, the error is small, with average deviations of 0.006. However, when the number of metallicity bins is reduced to 6, the deviation increases significantly, especially between 7.8 to 9.8 logarithmic age.

After that, we measured the uncertainties introduced by different isochrone models, Dartmouth, BaSTI, and PARSEC. Dartmouth provides a logarithmic age range between 8.4 and 10.15 with a total of 50 logarithmic age bins, and a metallicity range between -2.01 and 0.09 with a total of 22 metallicity bins. BaSTI provides a logarithmic age range between 7.48 and 10.15 with a total of 53 logarithmic age bins, and a metallicity range between -1.79 and 0.058 with a total of 8 metallicity bins. PARSEC provides a logarithmic age range between 6.6 and 10.15 with a total of 51 logarithmic age bins, and a metallicity range between -2.2 and 0.1 with a total of 24 metallicity bins.

%%%%%%%%%%
\begin{figure*}[ht!]
% \epsscale{1}
\plotone{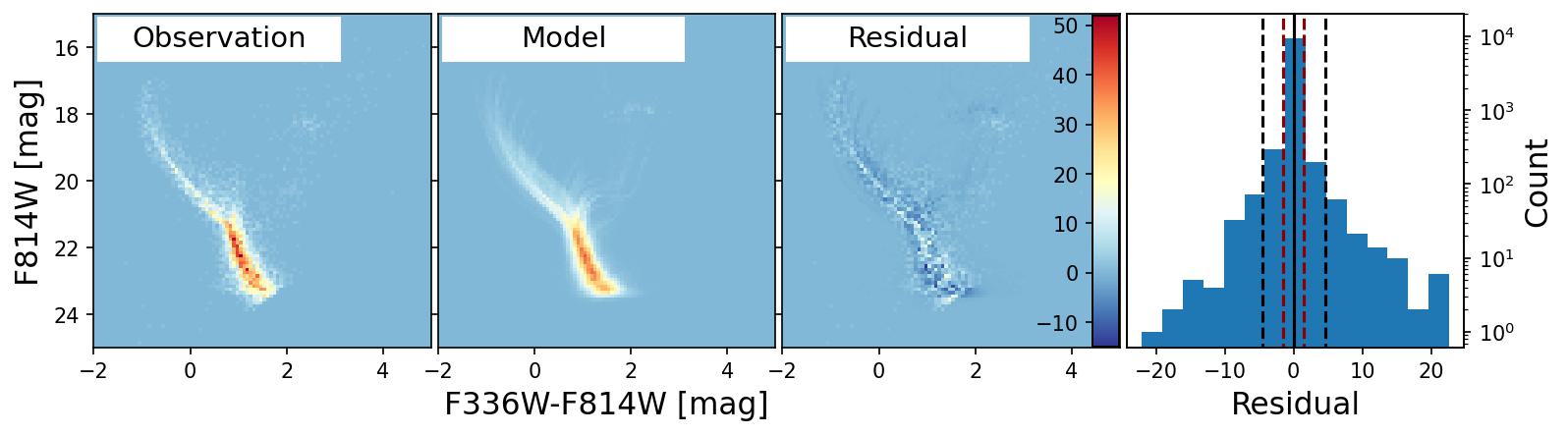}
\caption{The fitting result of NGC 1755. Three panels are the observation, model, and residual CMD intensity map with the color bar of star counts. residual = observation$-$model. The fourth panel shows the histogram of the residual. The black line shows the mean value of residual, which is $6.79\times10^{-17}$, while the red dashed lines show the STD ($1\sigma = 1.53$) and the black dashed lines show the $3\sigma = 4.58$.}
\label{fig:1755fit}
\vspace{10pt}
\end{figure*}
%%%%%%%%%%
%%%%%%%%%%
\begin{figure}[h!]
% \epsscale{0.9}
\plotone{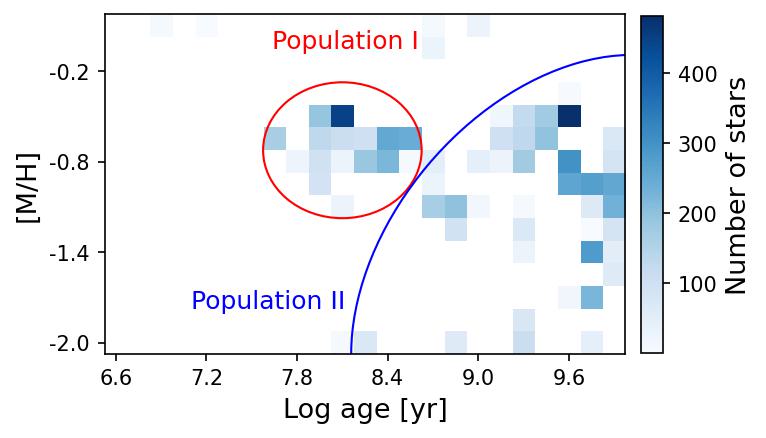}
\caption{The age-metallicity map of NGC 1755 with the color bar of the number of stars. The result shows two different populations, circled in red (population I) and blue (population II). The model CMDs of populations I and II are shown in the same color in Figure \ref{fig:1755cmd}.}
\label{fig:1755MAmap}
\end{figure}
%%%%%%%%%%
%%%%%%%%%%
\begin{figure}[h!]
% \epsscale{0.9}
\plotone{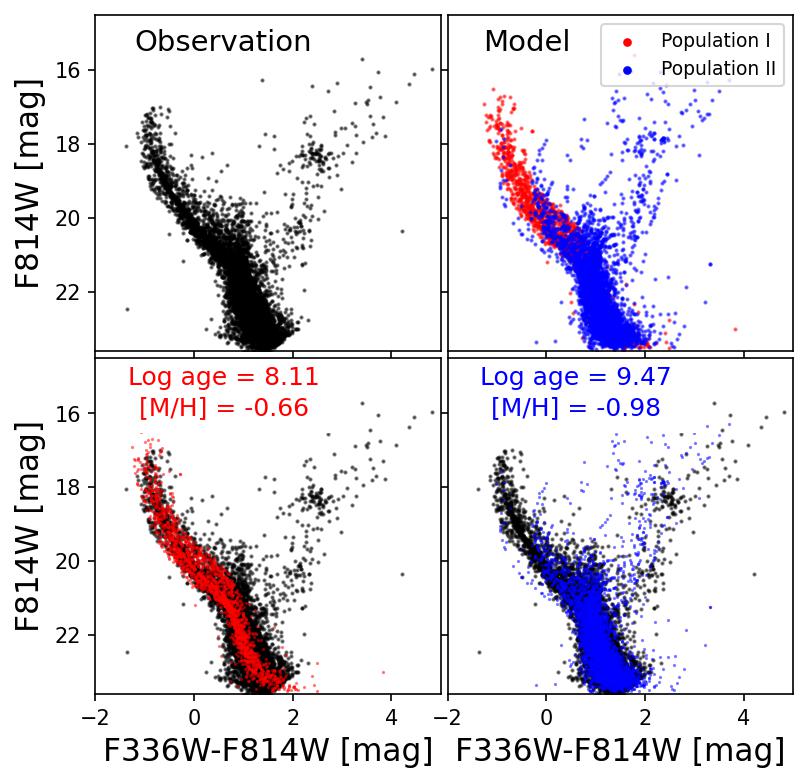}
\caption{The observation and model CMD of NGC 1755. The observation CMD points are in black, while the two populations of model CMD points are in red and blue, respectively. The bottom panels show the two populations with the labeled average logarithmic age and metallicity.}
\label{fig:1755cmd}
\end{figure}
%%%%%%%%%%

The results are shown in Figure \ref{fig:isochrones}. The left panel has logarithmic age on the x-axis, while the right panel has linear age in Gyr. The figure shows that the average $\Delta$ cumulative SFH is small but the scatter ($\sigma$) is large. Different isochrones lead to the largest systematic uncertainties in $\Delta$ cumulative SFH, ranging between -0.2 and 0.3. Specifically, the Dartmouth isochrone causes a decrease at logarithmic age 8.5 yr, while the BaSTI causes a decrease at logarithmic age 9.0 yr.

%%%%%%%%%%%%%%%%%%%%%%%%%%
\section{Application to star clusters} \label{sec:starcluster}
We apply \textsc{pancake} to star cluster NGC 1755. The photometry data, artificial star test, and input parameters (distance modulus $(M-m)_0 = 18.2$ and reddening $A_v = 0.34$) are all taken from \cite{yang2021spatial}. The raw data were observed with the Ultraviolet and Visual Channel of the Wide Field Camera 3 (UVIS/WFC3) on board the HST, and reduced by the WFC3 modules in the DOLPHOT package\footnote{http://americano.dolphinism.com/dolphot.}.

When generating the template, we use PARSEC isochrones with the logarithmic age bins from 6.6 to 9.6 with intervals of 0.15 and from 9.6 to 10.1 with intervals of 0.25, and the metallicity bins from -2 to 0.1 with intervals of 0.15. These bins are sparser than those for galaxies because star clusters usually have simpler stellar populations (one or two components), which allows us to build sparser nets. The IMF is Kroup and the binary fraction is 35\%. Quadtree binning with threshold 5 is selected as the gridding method. The foreground star CMD task is not applied. The foreground star CMD task provided by \textsc{pancake} does not apply very well to star clusters. First, we considered only the stellar models of the Milky Way as foreground stars. For star clusters, such as NGC 1755, there are also many foreground stars or field stars within its host galaxy, the Large Magellanic Cloud (LMC). Second, the best-fitting core radius of NGC 1755 is 7.83 arcsec and the half-mass radius is 10.32 arcsec \citep{yang2021spatial}. Under the foreground stars model, the number of foreground stars within the observing scale of a star cluster is very small. In contrast, the typical field of view for HST observations of galaxies is 202 arcsec. Therefore, the foreground star CMD removal task is more suitable for the galaxy sample.

Figure \ref{fig:1755fit} shows the fitting result of NGC 1755, including the CMD intensity map of the observation, model, residual (observation $-$ model), and the histogram of the residual. The residual values are all within $\pm$20 star counts and most are around 0. The mean value of residual is $6.79\times10^{-17}$, while the STD ($1\sigma$) is 1.53, indicating a good fit.

%%%%%%%%
\begin{figure*}[ht!]
% \epsscale{1.1}
\plotone{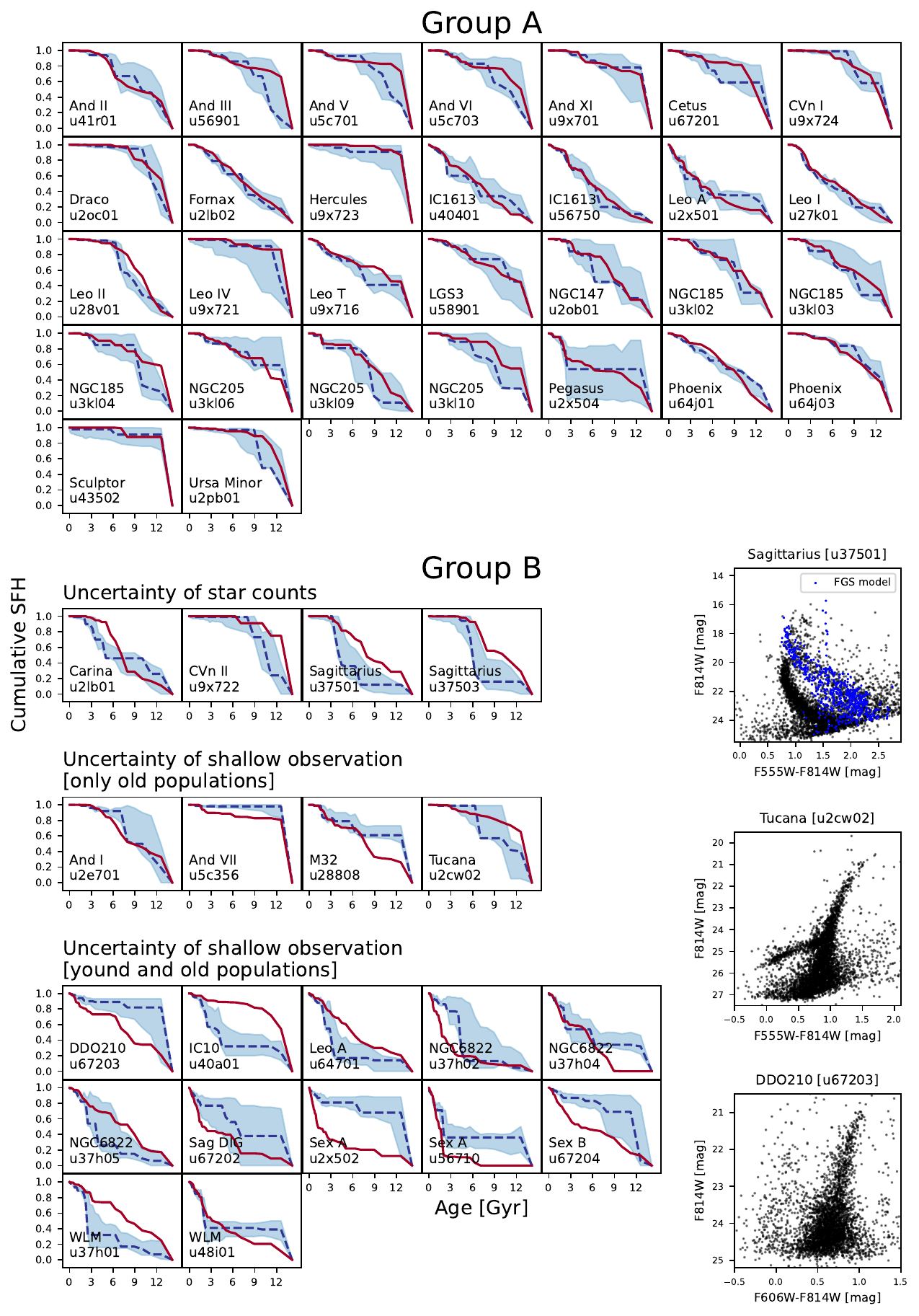}
\caption{The cumulative SFHs of dwarf galaxies, including \cite{weisz2014star1} in blue with random uncertainties and this paper in red. We divide the galaxies into Group A and Group B. Group A shows agreement between \cite{weisz2014star1} and this paper, while Group B shows offset. Group B is further divided into three classes, see text for details. For each class in Group B, an example of CMD distribution (black points) is given in the figure. Sagittarius[u37501] is a special example, which not only suffers from the lower star counts but is also strongly affected by foreground stars. For Sagittarius, the foreground stars model gives 850 star counts, already occupying 20.9\% of the total 4059 star counts.}
\label{fig:com_Weisz}
\end{figure*}
%%%%%%%%

Figure \ref{fig:1755MAmap} shows the age-metallicity map of NGC 1755, which presents two different populations. We compare the model CMDs of the two populations with the observed CMD in figure \ref{fig:1755cmd}. The population I in red, with the number of stars weighted average ${\rm log}\ (t/ \rm yr) = 8.11$ ($t \sim$ 128.8 Myr) and $\rm [M/H] = -0.62$, is the main population of star cluster NGC 1755. The average age is consistent with \cite{yang2021spatial}, who give an age of 127.0$\pm$87.12 Myr. The population II in blue, with the number of stars weighted average ${\rm log}\ (t/ \rm yr) = 9.47$ and $\rm [M/H] = -0.98$, is most likely foreground stars or field stars. Because we use the distance modulus of NGC 1755, which is not the real distance of foreground stars or field stars, the densest region of the asymptotic giant branch (AGB) stars for population II has a small offset between observed and model CMDs (see the lower-right panel in Figure \ref{fig:1755cmd}).

As a result, \textsc{pancake} is capable of accurately determining the age and metallicity of the star cluster.

%%%%%%%%%%%%%%
\section{Comparison with Weisz 2014} \label{sec:weisz}

\subsection{Data} \label{sec:weisz:data}
\cite{weisz2014star1} used \textsc{match} to analyze the SFHs of 40 Local Group dwarf galaxies, creating a publicly available SFH dataset using the CMD fitting method. This dataset is particularly advantageous for assessing the reliability of \textsc{pancake}. We use the same public photometry data as \cite{weisz2014star1} to avoid errors due to differences in photometry methods. The observational data used in this work were collected from the Local Group Stellar Photometry Archive\footnote{http://astronomy.nmsu.edu/logphot} \citep[LOGPHOT;][]{holtzman2006local}, which includes data from 38 nearby galaxies observed by the Hubble Space Telescope (HST) with the Wide Field Planetary Camera 2 (WFPC2). The data have been uniformly calibrated and reduced using HSTPHOT \citep{dolphin2000wfpc2}, which provides completeness information for the photometry. Most of the galaxies were observed with the filters F555W and F814W for more than 500 seconds. Others were observed with the filters F450W and F555W, or F606W and F814W. To balance data quality and the number of data points, we selected points with reliable magnitudes (\textit{magnitude} $<$ 90) and crowding less than 0.5 (\textit{crowding} $\leq$ 0.5) in both filters from the LOGPHOT survey output parameters to produce a high-quality catalog. The same criteria were applied for the selection of artificial stars. 

%%%%%%%%%%%%%%
\subsection{Results} 
We compare our fitting result of dwarf galaxies with that in \cite{weisz2014star1}. When generating the template CMDs, the input parameters are the same as in \cite{weisz2014star1}. The theoretical isochrone is PARSEC \citep{bressan2012parsec}. The logarithmic time has 51 bins within the range of $ 6.6 \leq {\rm log}\ (t/ \rm yr)\leq 8.7$ with intervals of 0.1, and of $ 8.7 \leq {\rm log}\ (t/ \rm yr)\leq 10.15$ with intervals of 0.05. The logarithmic metallicity has 24 bins within the range of $ -2.2 \leq {\rm [M/H]} \leq 0.1$ with intervals of 0.1. The IMF is Kroupa \citep{kroupa2001variation} and the binary fraction is $35\%$. The distance and extinction values are obtained from Table 4 in \cite{weisz2014star1}. Section \ref{sec:starcluster} shows that \textsc{pancake} can separate the suspected foreground or field stars in CMD, so we remove the significantly separated populations in age-metallicity maps of dwarf galaxies. 

The cumulative SFHs are presented in Figure \ref{fig:com_Weisz}, including \cite{weisz2014star1} and its random uncertainties in blue using the method of \textsc{match} \citep{dolphin2002numerical} and this paper in red. Although the quadtree binning with threshold 5 is suggested in section \ref{sec:gridtest}, here we use uniform binning with 100 bins in both color and magnitude axes to be consistent with \cite{weisz2014star1}. We divide galaxies into to Groups A and B. Group A is that the red falls into the blue region, which means that the results of \textsc{pancake} are consistent with \textsc{match} the expectation of random uncertainties. Group B is that the red falls out of the blue region, which means that some factors lead to significant systematic uncertainties.  Depending on the different factors, we further divide the Group B into three classes. For each class in Group B, Figure \ref{fig:com_Weisz} shows an example of the CMD distribution.

\begin{itemize}
\setlength{\itemsep}{-1pt}
    \item {\bf Group A:} Most galaxies compare well with median SFHs in \cite{weisz2014star1} and fall within the random uncertainty range (And II, And III, And V, And VI, And XI, Cetus, CVn I, Draco, Fornax, Hercules, IC 1613, Leo A [u2x501], Leo I, Leo II, Leo IV, Leo T, LGS 3, NGC 147, NGC 185, NGC 205, Pegasus, Phoenix, Sculptor, Ursa Minor).
    \item {\bf Group B - Class 1:} Some galaxies have the same trend as the SFH, but have a slight offset because of the lower star counts, which are less than 4000 star counts (Carina, CVn II, , Sagittarius). The example of the CMD distribution shown here is Sagittarius. Sagittarius not only suffers from the lower star counts, but is also strongly affected by foreground stars. For Sagittarius [u37501], the foreground stars model gives 850 star counts, already occupying 20.9\% of the total 4059 star counts.
    \item {\bf Group B - Class 2:} Some galaxies have very shallow observations and contain only old stellar populations that have not reached the Main Sequence Turn-off (MSTO), making it difficult to accurately fit their ages and metallicities. These galaxies show consistency in young populations, but offset in the middle-aged or old populations (And I, And VII, M32, Tucana).
    \item {\bf Group B - Class 3:} Some galaxies have very shallow observations and contain both young and old stellar populations, making the fitting condition more complex. These galaxies have the largest discrepancy compared to \cite{weisz2014star1}, and the offset is large in all populations (DDO 210, IC 10, Leo A [u64701], NGC 6822, Sag DIG, Sex A, Sex B, WLM).
\end{itemize}

Under the same assumptions (theoretical isochrone, IMF, binary fraction, distance, and extinction), the systematic uncertainty for gridding methods and the uncertainty for star counts will lead to an offset in the SFHs. In addition, the MSTO regions are extremely important in constraining the SFHs of galaxies. In the following, we give an example of a well-fitted galaxy, Fornax, which has deep observation and sufficient star counts. We also give an example of a poorly-fitted galaxy, DDO 210, because of the lack of MSTO information in the HST/WFPC2 observation.

%%%%%%%%%%%%%%
\subsection{Data with deep observation: Fornax} 
\begin{figure}[h!]
% \epsscale{0.9}
\plotone{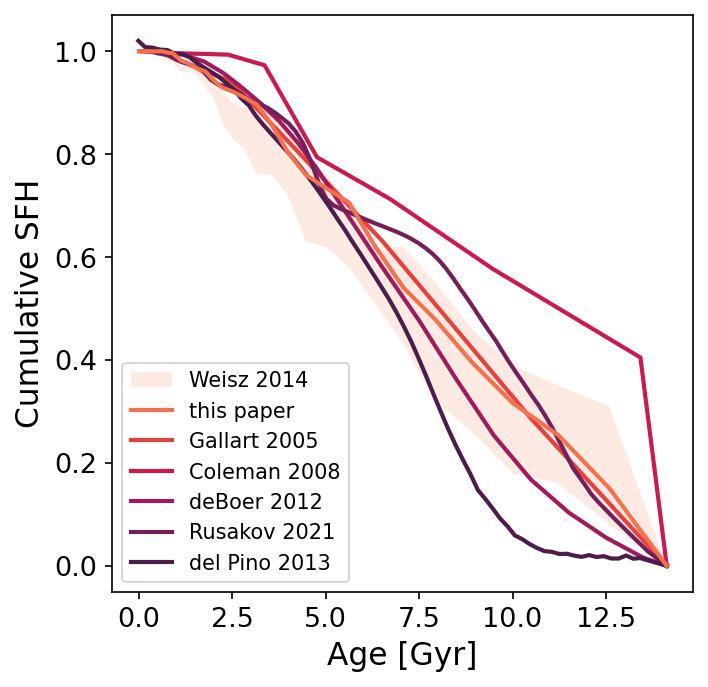}
\caption{The cumulative SFHs of Fornax. As in Figure \ref{fig:com_Weisz}, \cite{weisz2014star1} is presented in the random uncertainty error region, while the other works are shown as lines in different colors labeled in the figure. The reference, observation information, and CMD method input parameters of all the results are listed in Table \ref{tab:fornax1} and \ref{tab:fornax2}.}
\label{fig:fornax}
\end{figure}

The galaxies that have deep observations and MSTO information can be fitted well. We use the well-studied Fornax dwarf galaxy as an example. We compare the SFH through CMD methods using different observation data and different fitting details, shown in Figure \ref{fig:fornax}. 
% Some literature cumulative SFHs are derived from SFR $M_{\odot} yr^{-1}$ in SFH, which have a more sparse in the timebin (shown as dots and lines in the figure). For those cumulative SFHs directly reported in the literature, only the line is shown in the figure. 
Despite the various observational telescopes, filters, fields, and fitting methods, the trends of cumulative SFHs are consistent. Observation information and CMD method details are listed in Tables \ref{tab:fornax1} and \ref{tab:fornax2}, respectively. 

%%%%%%%%%%
%%%%%%%%%%

The result of this paper is consistent with \cite{weisz2014star1}, utilizing the same data, theoretical isochrone, and CMD method input parameters. The result of \cite{gallart2005star} is closer to \cite{weisz2014star1} and this paper with the same isochrone library, while the field of the WFPC2 observation is within the field of VLT-FORS1.

\cite{gallart2005star} and \cite{del2013spatial} both use the V, I bands from VLT-FORS1 in July 2000, and used \textsc{IAC} to build the model CMDs. However, they use different theoretical isochrone libraries and fitting methods, which may cause the difference in the old population. \cite{del2013spatial} presented more middle-aged populations (7-10 Gyr) and less old populations (10-13.5 Gyr) in the central region of Fornax. 

\cite{rusakov2021bursty} and \cite{del2013spatial} show large differences in the trend of cumulative SFHs between 6-14 Gyr, despite using the same isochrone library and similar fitting parameters. The field of \cite{rusakov2021bursty} is located in the northeast direction of the Fornax galactic center, while the field of \cite{del2013spatial} is located in the southwest direction (see Tabel \ref{tab:fornax1}). This may imply that the populations around the Fornax center are not symmetric. 

\cite{coleman2008deep} and \cite{de2012star} both use a large field-of-view observations that include the whole information of Fornax within a specific elliptical radius. Although they use different methods, on average \cite{coleman2008deep}, which has a larger field, has more old populations than \cite{de2012star}.

\begin{deluxetable*}{cccccccccc}[ht!]
\tablenum{2}
\tablecaption{Observation information of Fornax dwarf galaxy \label{tab:fornax1}}  
\tablewidth{0pt}
\tablehead{
\colhead{References} & \colhead{Instruments$^a$} &  \colhead{filters} &  \colhead{observed fields} & \colhead{G radius [']} & \colhead{star counts} 
}
\startdata
\cite{gallart2005star} &  VLT-FORS1  & V , I       & 408" $\times$ 408" & 1.2 & 82,000\\
\cite{coleman2008deep} &  ESO/MPG-WFI& B , R       & $\rm r_{ell}< 76'$ & 0.0 & 151,000\\
\cite{de2012star}      &  CTIO-MOSAIC& B , V       & $\rm r_{ell}< 48'$  & 0.0 & 270,000\\
\cite{del2013spatial}  &  VLT-FORS1  & V , I       & 408" $\times$ 408" & 1.2 & 69,590\\
\cite{weisz2014star1}  &  HST-WFPC2  & F555W,F814W & $\sim$160" $\times$ 160" & 3.5 & 18620\\
\cite{rusakov2021bursty}& HST-ACS/WFC &F475W,F814W & 202" $\times$ 202" & 2.7 & 41,656\\
This paper             &  HST-WFPC2  & F555W,F814W & $\sim$160" $\times$ 160" & 3.5 & 18620\\
\enddata
\tablecomments{
\begin{itemize}
    \item[a] Instruments name:\\
    VLT-FORS1: FOcal Reducer and low dispersion Spectrograph for the Very Large Telescope,\\ 
    ESO/MPG-WFI: ESO/MPG 2.2 m telescope equipped with the Wide Field Imager,\\
    CTIO-MOSAIC: CTIO 4-m MOSAIC II camera,\\
    HST-ACS/WFC: Advanced Camera for Surveys Wide Field Channel for the Hubble Space Telescope;
    \item[b] $\rm r_{ell}$ is the elliptical radius of Fornax;
    \item[c] G radius is the distance from the field center to the Fornax galaxy center.
\end{itemize}
}
\end{deluxetable*}

\begin{deluxetable*}{cccccccccc}[ht!]
% \vspace{-1cm}
\tablenum{3}
\tablecaption{CMD fitting parameters of Fornax dwarf galaxy \label{tab:fornax2}}  
\tablewidth{0pt}
\tablehead{
\colhead{References} & \colhead{CMD Method$^a$} &
\colhead{Isochrones$^b$} & \colhead{$\rm age_{max}^c$} & \colhead{IMF$^d$} & \colhead{$f^e$} & \colhead{(M-m)$_0$} & \colhead{A$_v$} \\
\colhead{} & \colhead{} &\colhead{} & \colhead{[Gyr]} & \colhead{} & \colhead{[\%]} & \colhead{} & \colhead{} 
}
\startdata
\cite{gallart2005star} & \textsc{IAC} & PARSEC & $\sim$13 & Kroupa & 10-90 & - & -\\
\cite{coleman2008deep} & \textsc{match} & PARSEC & 15.85 & Salpeter & 50 & 20.84 & 0.047-0.093\\
\cite{de2012star} & Talos & Dartmouth & 12.5 & - & - & 20.70 & 0.093\\
\cite{del2013spatial} & \textsc{IAC} & BaSTI & 13.5 & Kroupa & 40 & 20.66 & 0.068\\
\cite{weisz2014star1}  & \textsc{match} & PARSEC & 12.58 & Kroupa & 35 & 20.79 & 0.06\\
\cite{rusakov2021bursty} & method by Rusakov & BaSTI &13.9&Kroupa &50&20.818 & 0.068\\
This paper    & \textsc{pancake} & PARSEC & 14.13 & Kroupa & 35 & 20.79 & 0.06\\
\enddata
\tablecomments{
\begin{itemize}
    \item[a] CMD fitting reference, \textsc{match} \citep{dolphin2002numerical}, \textsc{IAC} \citep{aparicio2009iac}, Talos \citep{de2012star}, method by Rusakov \citep{rusakov2021bursty}; 
    \item[b] theoretical isochrones reference, PARSEC \citep{bressan2012parsec}, Dartmouth \citep{dotter2008dartmouth}, BaSTI \citep{pietrinferni2004large};
    \item[c] $\rm age_{max}$ is maximum age bin;
    \item[d] IMF reference, Kroupa \citep{kroupa2001variation}; Salpeter \citep{salpeter1955luminosity};
    \item[e] $f$ is the binary fraction.    
\end{itemize}
}
\end{deluxetable*}

%%%%%%%%%
\begin{figure}[h!]
% \epsscale{0.9}
\plotone{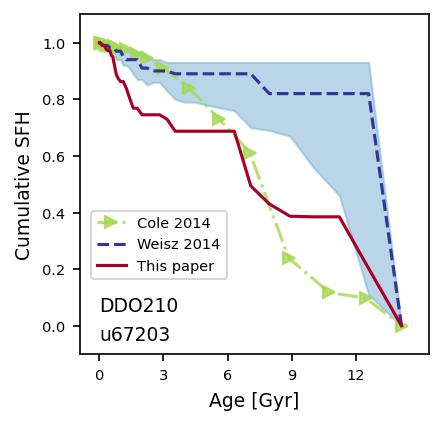}
\caption{The cumulative SFHs of DDO 210 galaxies. Same as Figure \ref{fig:fornax}, \cite{weisz2014star1} is presented in blue with an error region and this paper is in red. \cite{cole2014delayed} is in green with dots. \cite{cole2014delayed} used a deeper observation from HST-ACS/WFC that contains the main sequence region of DDO 210.}
\label{fig:ddo210}
\end{figure}
%%%%%%%%%

\subsection{Data with shallow observation: DDO 210} \label{sec:ddo210}
The galaxies with shallow observations without MSTO information do not fit well. We use the DDO 210 dwarf galaxies as an example. We compare the SFH obtained through a deeper observation and contain the main sequence region of DDO 210 \citep{cole2014delayed}. \cite{cole2014delayed} use the F475W and F814W bands in HST-ACS/WFC observation, leaving a sample of 51,239 star counts. The CMD fitting method is \textsc{Cole07} using the PARSEC isochrone with the maximum age bin of 13.18 Gyr. They use the Chabrier IMF \citep{chabrier2003galactic} and 65\% binary fraction, in which 25\% of the binaries are defined as “close” binaries with the secondary mass drawn from a flat IMF. Figure \ref{fig:ddo210} shows the cumulative SFHs of DDO 210. The main sequence stars balance the ratio among the young, middle-aged, and old populations. \cite{cole2014delayed} presents only one clear starburst around 6 Gyr. Without the main sequence, \cite{weisz2014star1} and this paper both show starbursts in the middle-aged population around 6 Gyr and in the old population around 12 Gyr. Differently, this paper shows a starburst in the young population around 1 Gyr (see also Figure \ref{fig:appendix_pancake_results}).

%%%%%%%%%%%%%%%%%
\section{Comparison with FUV SFR} \label{sec:FUV}
\begin{figure}[h!]
% \epsscale{0.9}
\plotone{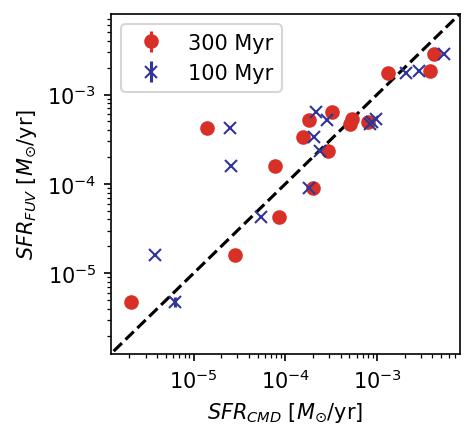}
\caption{The comparison between FUV SFR and CMD SFR in 100 Myr (blue crosses) and 300 Myr (red dots). The black dashed line represents the 1:1 line.}
\label{fig:FUV}
\end{figure}

FUV is a good tracer of the recent star-forming process in galaxies. The FUV data are obtained from the Galaxy Evolution Explorer \citep[GALEX;][]{martin2005galaxy}, which provides near-ultraviolet (NUV) and FUV images. Each image has a pixel size of 1.5 arcsec. The typical full width at half-maximum of the point spread function is 5.0 arcsec for FUV and 5.5 arcsec for NUV. The images are from the Nearby Galaxy Survey (NGS), Medium Imaging Survey (MIS), Deep Imaging Survey (DIS), Guest Investigator Program (GII), and Exceptional Imaging Survey (ETI). There are 19 galaxies covered in the GALEX survey with both FUV and NUV images. After photometry, we remove samples with CMD or FUV SFR less than $10^{-6} M_{\odot} \rm \ yr^{-1}$, as they have huge detection errors. Finally, Figure \ref{fig:FUV} includes 10 galaxies (16 fields). All the {\it GALEX} data used in this paper are included in the GALEX Merged Catalog of Sources (MCAT) \citep{https://doi.org/10.17909/t9h59d}, which can be found in MAST: \dataset[10.17909/T9H59D]{http://dx.doi.org/10.17909/T9H59D}.

%%%%%%%%%%%%%%%%%
%%%%%%%%%%%%%%%%%

FUV traces the SFR within the previous 100-300 Myr \citep{kennicutt2012star}. We compare FUV SFR with the CMD SFR within 100 Myr (blue crosses) and 300 Myr (red dots). As expected, the field with a larger SFR has a smaller scatter between CMD and FUV SFR. The normalized root-mean-square deviation ($\frac{1}{(y_{\rm max}-y_{\rm min})}\sqrt{\frac{\sum (x-y)^2)}{n}}$) of 100 Myr and 300 Myr in Figure \ref{fig:FUV} are 0.25 and 0.22, respectively. 

%%%%%%%%%%%%%%%%%
\subsection{GALEX photometry}
We use a method similar to \cite{wang2017local} for the GALEX photometry, which relies on the Source Extractor \citep{bertin1996sextractor}. However, instead of the growth curve and total flux of galaxies in \cite{wang2017local}, we use FUV SFR maps of the dwarf sample to select the same field of view of WFPC2 and compare with CMD SFR. 

First, we use Source Extractor to produce a segmentation map for each image, to remove the global background and neighboring sources. Second, we select pixels with flux ratios NUV/FUV $>$ 5 and NUV signal-to-noise ratio (S/N) $>$ 2, to remove the foreground stars. Third, we measure the radial profile of galaxies and identify the radius where the profile flattens out in the noise, to remove the residual background. Finally, we use the function in \cite{murphy2011calibrating} to transform the FUV luminosity to SFR,
\begin{equation}
    \left(\frac{\rm SFR_{FUV}}{M_{\odot} \rm  \ yr^{-1}}\right) = 4.42 \times 10^{-44} \left(\frac{\rm L_{FUV}}{\rm erg \ s^{-1}}\right).
\end{equation}
After obtaining the FUV SFR map, we select the WFPC2 field center and cut a circle with a radius of 80" to obtain the relative FUV SFR of each WFPC2 field.

\section{Limitations} \label{sec:limitation}
The foreground stars can affect the accuracy of the \textsc{pancake}. On the scale of star clusters, the stellar population parameters of most likely foreground star components can be clearly distinguished from that of the target cluster, as described in Section \ref{sec:starcluster}. This clear distinction allows for precise modeling and analysis within individual star clusters. On a galactic scale, however, the influence of foreground stars becomes more pronounced. The fundamental assumption of \textsc{pancake} is that stars are formed in star clusters with uniform metallicity and at the same epoch. Foreground stars, originating from various star formation events and possessing diverse metallicities, may not conform to this assumption. To mitigate the effects of foreground stars in galaxies, we generate a Milky Way stellar population model based on the stellar structure \citep{de2010mapping}. This model effectively removes most of the foreground population in CMDs, although it has limitations as discussed in Section \ref{sec:CMDmethod:foreground}. The best solution for foreground stars is to select a comparable field region during observations, where the accuracy of foreground and field stars can be measured through photometry.

Blending refers to images with such a high density of detected sources that distinguishing individual objects becomes difficult. Under a good quantity of observation, the effects of blending are considered in both photometry and fake star detection, through the crowding and completeness parameters. However, we do not provide an improved solution for data that is limited by the target distance or shallow photometry data, as discussed in Section \ref{sec:weisz} that the MSTO information influences the fitting. 

\vspace{1cm}

\section{Summary}\label{sec:summary}
We develop a python-based open-source CMD fitting package, \textsc{pancake}, which contains tasks for preprocessing, template CMDs generation, foreground star CMD generation and removal, CMD gridding, and CMD fitting. The input is highly autonomous, i.e. compatible with all theoretical isochrones in the specified format, providing three optional IMFs and three optional gridding methods. \textsc{pancake} provides accurate stellar population parameters through an efficient model-free \textsc{PyTorch} gradient descent CMD fitting. 

We use mock data to test the systematic uncertainties for gridding methods, IMFs, binary fractions, star counts, and isochrones. We define a normalized $\chi^2$ to quantify the uncertainties. We suggest the quadtree binning with a threshold of 5. The IMF and binary fraction of the target galaxy may be derived using the Monte Carlo method by running the code multiple times. We also describe the uncertainties by $\Delta$ cumulative SFH. The binary fraction affects the stellar population mostly between 0-8 Gyr, while the IMF affects the population between 7-12 Gyr. Isochrones are the most significant effects for $\Delta$ cumulative SFH. Binary fractions and star counts are two other important factors affecting $\Delta$ cumulative SFH.

We then apply \textsc{pancake} to a star cluster. \textsc{pancake} can accurately obtain the stellar population parameters of the star cluster and compare well with the literature. We also apply \textsc{pancake} to 38 dwarf galaxies (50 fields) observed by WFPC2/HST and compare the results with those of \cite{weisz2014star1}, who use \textsc{match} under the same situation. Most of the galaxies are consistent well and fall inside the error region of \cite{weisz2014star1}. Part of the galaxies fall outside the error region, because of significant uncertainty for star counts or observational limitations. We compare the SFRs derived by \textsc{pancake} with those obtained by FUV measurements. The results of these two methods are consistent, validating the accuracy and reliability of \textsc{pancake} in determining recent SFRs of nearby galaxies.

We have integrated isochrone libraries generated using PARSEC for most optical and near-infrared filters commonly used in space telescopes such as HST, JWST, and the upcoming CSST. This integration ensures compatibility and accuracy when analyzing data from these instruments. However, users still have the option to select other isochrone libraries according to their preferences and specific requirements.

\textsc{pancake} is able to be integrated as an optional tool in the CSST data server at Zhejiang Lab, enhancing its accessibility and usability for researchers in the field.

%%%%%%%%%%%%%%%%%%%%%%%%%%%
\begin{acknowledgments}
This work is supported by NSFC grants No. 11988101; National Key R\&D Programs of China No. 2023YFC2206403 and 2024YFA1611602; China Manned Space Program with grant No. CMS-CSST-2025-A08; and NSFC grants No. 12373012, 12041302, and 12203064. DL is a new cornerstone investigator.
\end{acknowledgments}

\vspace{5mm}
% \facilities{HST(STIS), Swift(XRT and UVOT), AAVSO, CTIO:1.3m,
% CTIO:1.5m,CXO}

%% Similar to \facility{}, there is the optional \software command to allow 
%% authors a place to specify which programs were used during the creation of 
%% the manuscript. Authors should list each code and include either a
%% citation or url to the code inside ()s when available.

\software{NumPy \citep{harris2020array},        
          Astropy \citep{astropy:2013, astropy:2018, astropy:2022},
          Scipy \citep{2020SciPy-NMeth},
          qthist2d\footnote{https://github.com/jradavenport/qthist2d},
          \textsc{PyTorch} \citep{paszke2019pytorch},
          Matplotlib \citep{Hunter:2007},
          pandas \citep{reback2020pandas},
          extinction\footnote{https://github.com/sncosmo/extinction},
          Source Extractor \citep{bertin1996sextractor}
          }

%% Appendix material should be preceded with a single \appendix command.
%% There should be a \section command for each appendix. Mark appendix
%% subsections with the same markup you use in the main body of the paper.

%% Each Appendix (indicated with \section) will be lettered A, B, C, etc.
%% The equation counter will reset when it encounters the \appendix
%% command and will number appendix equations (A1), (A2), etc. The
%% Figure and Table counter will not reset.

\appendix
\section{Data compilation - PANCAKE $\&$ GALEX photometry files}\label{appendix_data}
\begin{figure*}[ht!]
% \epsscale{0.9}
\plotone{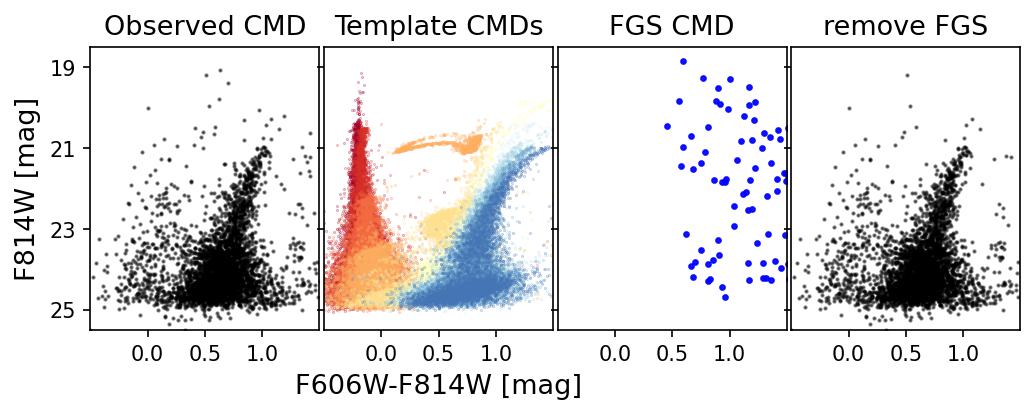}
\plotone{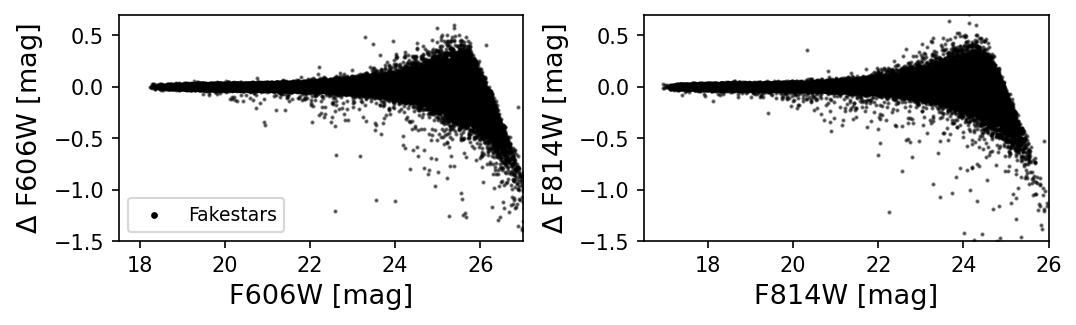}
\plotone{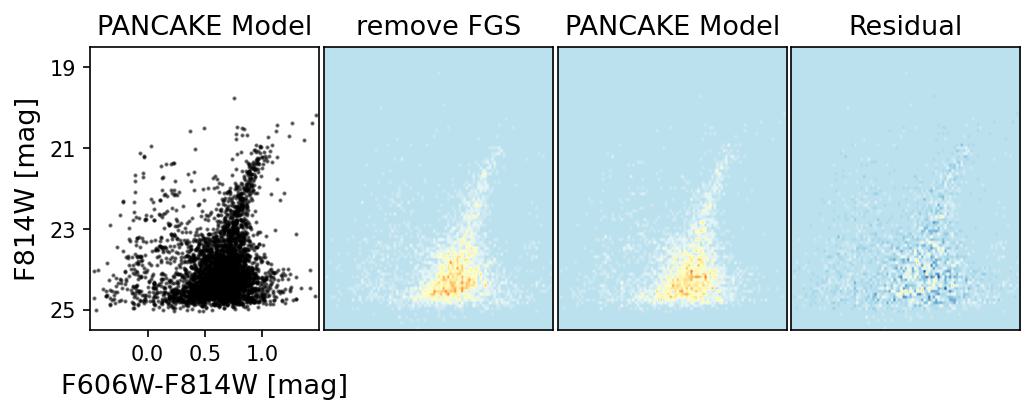}
\caption{Atlas of files during the processing of \textsc{pancake}, using DDO 210 dwarf as an example. The first row shows the observed CMD, examples of the template CMDs, the foreground stars CMD, and the observed CMD after foreground stars removal; The second row shows the artificial stars results of two filters; The third row shows the \textsc{pancake} model CMD and the 2D histogram figures for the observed CMD after foreground stars removal, the \textsc{pancake} model CMD, and the residual CMD (observed - model).}
\label{fig:appendix_pancake_process}
\end{figure*}

\begin{figure*}[ht!]
% \epsscale{0.9}
\plotone{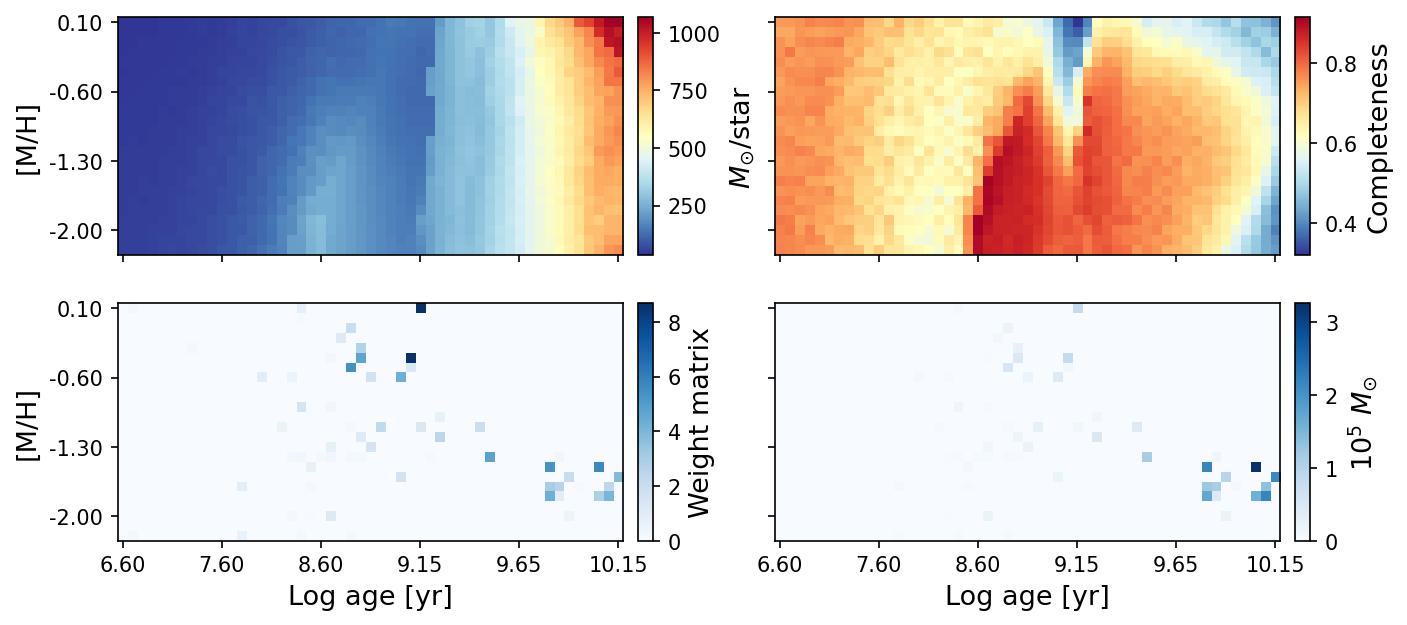}
\plotone{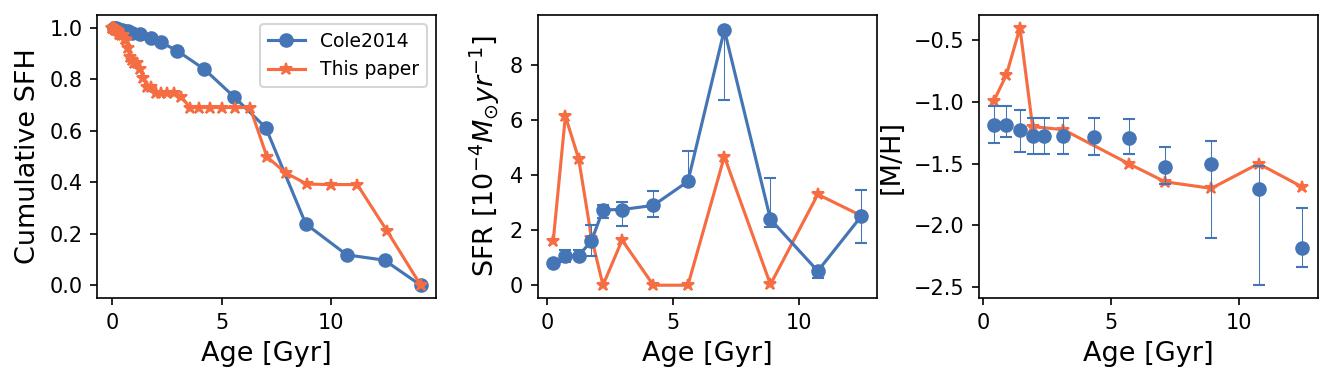}
\caption{Atlas of files during the transformation of the \textsc{pancake} fitting results, using DDO 210 dwarf as an example. The first row shows the number of stars to mass ratio map (SFM-to-$\bf S$) and the observational completeness map ($\bf C$); The second row shows the fitting output weight matrix $\bf W$ and the output age-metallicity map in mass ($\bf SFM$); The third row shows the cumulative SFH, SFR, and age-metallicity relation, compared with \cite{cole2014delayed}, see discussion in Section \ref{sec:ddo210}.}
\label{fig:appendix_pancake_results}
\end{figure*}

\begin{figure*}[ht!]
% \epsscale{0.9}
\plotone{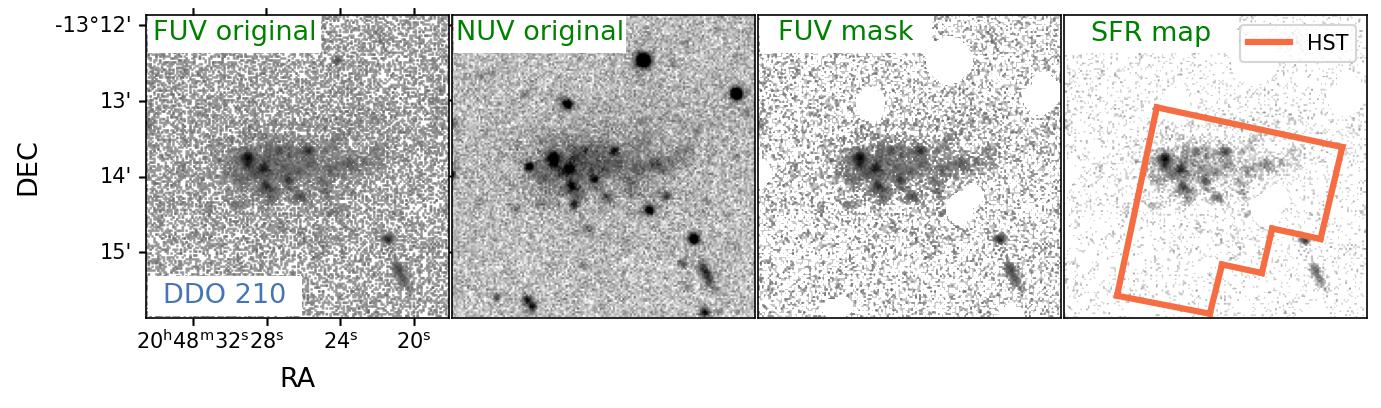}
% \plotone{./figures/Appendix_galex2.jpg}
\caption{Atlas of files during the GALEX photometry, using DDO 210 dwarf as an example. The four panels show the cut FUV and NUV original data, the FUV data after background removal and masking, and the SFR map. The orange region in the fourth panel shows the WFPC2/HST field of view.}
\label{fig:appendix_galex}
\end{figure*}

We release all of the input data, intermediate steps, and output results of this work in Science BD \citep{scienceBDcode}, including the \textsc{pancake} files for star cluster NGC 1755 and 38 dwarf galaxies (50 fields), and the GALEX photometry files. These files are listed in Table \ref{tab:data_release}. Using DDO 210 dwarf as an example, Figures \ref{fig:appendix_pancake_process} and \ref{fig:appendix_pancake_results} present files of \textsc{pancake} process and Figure \ref{fig:appendix_galex} presents the GALEX photometry files. 

Figure \ref{fig:appendix_pancake_process} presents the processing of \textsc{pancake}. The first row shows the observed CMD in black, several examples of the template CMDs in different colors, the foreground stars CMD in blue, and the observed CMD after foreground stars removal in black; The second row shows the artificial stars results of two filters; The third row shows the \textsc{pancake} model CMD result in black and the comparison between observed and model CMD in 2D histogram figures (residual = observed - model).

Figure \ref{fig:appendix_pancake_results} shows the transformation of the \textsc{pancake} fitting results. The first row shows the number of stars to mass ratio map (SFM-to-$\bf S$) and the observational completeness map (matrix $\bf C$); The second row shows the fitting output weight matrix $\bf W$ and the output age-metallicity map ($\bf SFM$) in mass; The third row shows the cumulative SFH, SFR, and age-metallicity relation. We compare with \citep{cole2014delayed} here, as \cite{weisz2014star1} only supports cumulative SFH which cannot be used to derive the SFR and age-metallicity relation.

Figure \ref{fig:appendix_pancake_results} shows the GALEX photometry process. The four panels show the cut FUV and NUV original data, the FUV data after background removal and masking, and the SFR map. The orange region in the fourth panel shows the WFPC2/HST field of view.

\section{PANCAKE-specific input format}\label{appendix_format}

Before running \textsc{pancake}, we need a preprocessing task to transform the observational data and theoretical isochrones (green boxes in Figure \ref{fig:workflow}) into the \textsc{pancake}-specific format.

The observational data include photometry information for true stars and the results of all artificial stars tests. Using the LOGPHOT catalog as an example, to balance data quality and the number of data points, we selected points with reliable magnitudes (\textit{magnitude} $<$ 90) and crowding less than 0.5 (\textit{crowding} $\leq$ 0.5) in both filters. The output should be two NumPy arrays of x-axis (color) and y-axis (magnitude) values for each of the data points selected in CMD. The output format are `npy' files named by `target$\_$obs$\_$cmdx.npy' and `target$\_$obs$\_$cmdy.npy', respectively, where the `target' could be changed to the name of the target galaxy or any other name you like. The same criteria were applied for the selection of artificial stars. The output should be four NumPy arrays of input star magnitude values and output star magnitude values after the photometry process of two bands. The output format are `npy' files named by `target$\_$fakestar$\_$input$\_$filter1.npy', `target$\_$fakestar$\_$output$\_$filter1.npy', `target$\_$fakestar$\_$input$\_$filter2.npy', and `target$\_$fakestar$\_$output$\_$filter2.npy'.

Theoretical isochrones need to be cut into single isochrone tables in `csv' format. For example, the name of the table should be `M-0.0500A6.6000', where -0.0500 is the metallicity and 6.6000 is the age in logarithms. The metallicity and age numbers are the relative values of the isochrone. Numbers must be stored to 4 decimal places, regardless of positive or negative. For each isochrone table, we need `Mini', `Mass', `F$\_$filter1$\_$Wmag'(for example `F475Wmag'), `F$\_$filter2$\_$Wmag'(for example `F814Wmag'), `label'. `Mini' is the IMF mass in $M_{\odot}$ and `Mass' is the actual mass in $M_{\odot}$ after considering the mass loss during the stellar evolution. `F475Wmag' and `F814Wmag' are the two-bands magnitudes. `label' is the evolutionary stage. The label of main-sequence stars is equal to 1. Theoretical isochrones contain $I = t \times z$ number of isochrones, depending on the age bins $t$ and metallicity bins $z$ set for the theoretical stellar evolutional model. For reading the isochrone tables, the age bins $t$ and metallicity bins $z$ should be saved as two NumPy arrays in `npy' files named by `logage$\_$list.npy' and `metal$\_$list.npy', respectively. All the tables and two NumPy arrays of age and metallicity should be saved in one folder named `newstore'.

\begin{deluxetable*}{ll}[ht!]
\tablenum{4}
\tablecaption{Data release file list \label{tab:data_release}}  
\tablewidth{0pt}
\tablehead{
\colhead{filename} & \colhead{file content}
}
\startdata
target$\_$obs$\_$cmdx(y).npy & \textsc{pancake} -- The observed CMDs in the form of \textsc{pancake}\\
target$\_$fakestar$\_$input(output)$\_$filter.npy & \textsc{pancake} -- The artificial stars in the form of \textsc{pancake}\\
template$\_$target$\_$filter.npy & \textsc{pancake} -- The template CMDs (two-band magnitudes)\\
target$\_$FGS$\_$x(y).npy & \textsc{pancake} -- The foreground stars CMDs\\
target$\_$obs$\_$cmdx(y)$\_$removeFGS.npy & \textsc{pancake} -- The observed CMDs removed foreground stars\\
number$\_$to$\_$IMF(ACT)$\_$mass$\_$target.npy & \textsc{pancake} -- The number of stars to mass ratio SFM-to-$\bf S$\\
template$\_$target$\_$completeness.npy & \textsc{pancake} -- The observational completeness matrix $\bf C$ \\
target$\_$cmd.pdf & \textsc{pancake} -- The figures of observed, model, and residual CMDs\\
target$\_$fitres.npy & \textsc{pancake} -- The fitting output weight matrix $\bf W$ \\
target$\_$sfh$\_$msun.npy & \textsc{pancake} -- The output age-metallicity map in mass $\bf SFM$ \\
target$\_$f(n)cut($\_$smooth).fits & GALEX photometry -- Cut FUV/NUV original data \\
target$\_$f(n)cut$\_$CB.fits & GALEX photometry -- Background remove \\
target$\_$f(n)cut$\_$CB$\_$PB.fits & GALEX photometry -- Profile background remove \\
target$\_$f(n)cut$\_$M1($\_$CB$\_$PB$\_$MS).fits & GALEX photometry -- Mask \\
target$\_$SFR(SFRsurface)($\_$err).fits & GALEX photometry -- SFR map and error \\
\enddata
\tablecomments{
The `target' in the filename needs to be replaced with the specific name of the galaxy or star cluster, such as `ddo210'. The `filter' in the filename needs to be replaced with the specific band, such as `555'/`814'.}
\end{deluxetable*}

\bibliography{sample631}{}
\bibliographystyle{aasjournal}

%% This command is needed to show the entire author+affiliation list when
%% the collaboration and author truncation commands are used.  It has to
%% go at the end of the manuscript.
%\allauthors

%% Include this line if you are using the \added, \replaced, \deleted
%% commands to see a summary list of all changes at the end of the article.
%\listofchanges

\end{document}